\documentclass[11pt,a4paper]{article}
\usepackage[english]{babel}
\usepackage{cite}

\usepackage{jheppub}
\usepackage{epsf}
\usepackage{amssymb}
\usepackage{amsmath}
\usepackage{amsfonts}
\usepackage{psfrag,epsfig,graphicx,graphics}
\usepackage{subcaption}
\usepackage{wasysym}
\usepackage{setspace}

\title{Low $x$ physics as an infinite twist (G)TMD framework: unravelling the origins of saturation}
\author[a]{Tolga Altinoluk}
\author[b]{and Renaud Boussarie}

\affiliation[a]{National Centre for Nuclear Research, 
00-681 Warsaw, Poland}
\affiliation[b]{Physics Department, Brookhaven National Laboratory, 
Upton, NY 11973, USA}

\emailAdd{tolga.altinoluk@ncbj.gov.pl}
\emailAdd{rboussarie@bnl.gov}

\abstract{We show how the formulations of low $x$ physics involving Wilson line operators can be fully rewritten into an infinite twist TMD or GTMD framework, respectively for inclusive and exclusive observables. This leads to a perfect match between low $x$ physics and moderate $x$ formulations of QCD in terms of GTMDs, TMDs, GPDs or PDFs. We derive the BFKL limit as a kinematic limit and argue that beyond the Wandzura-Wilczek approximation, 3-body and 4-body unintegrated PDFs should be taken into account even in this regime. Finally, we analyze how saturation should be understood as three distinct effects: saturation through non-linearities in the evolution equations at small $x$, saturation through multiple interactions with slow gluons as TMD gauge links, and saturation as the enhancement of genuine twist corrections.}

\begin{document}

\pagestyle{empty} \newpage{}

\mbox{%
}

\maketitle

\pagestyle{plain}

\setcounter{page}{1}


\section{Introduction}

The continuity between moderate $x$ and low $x$ observables has
been the subject of many studies in perturbative QCD. The factorization
schemes involved in both cases seem indeed very different at first
sight. For processes where the center-of-mass energy $s$ is of the
same order as the hard partonic scale $Q^{2}$, i.e. at moderate $x=Q^{2}/s$,
and for inclusive enough observables, colinear factorization applies.
The simplicity of having a single hard scale carried by a single hard
momentum allows for a standard Operator Product Expansion
(OPE) to be performed. Such an OPE consists in the expansion of a bilocal operator into a discrete set of local operators $\mathcal{O}_n$, usually ordered according to their twist (dimension $-$ spin). For example for quark currents $J$ the OPE has the form:
\begin{equation}
J(z)J(0)\rightarrow\sum_{n}C_{n}\left(z,\mu\right)\mathcal{O}_{n}\left(\mu\right),\label{eq:colOPE}
\end{equation}
where $\mu$ is a renormalization scale and where divergences from
the loop corrections to the Wilson coefficient $C_{n}\left(z,\mu\right)$
are cancelled via the renormalization of the local operator $\mathcal{O}_{n}\left(\mu\right)$.
This renormalization allows to resum large logarithms of the hard
scale $Q$. 
In the low $x$ limit, the hardest scale of the process is given by
$s$, and the previous OPE is not convenient to address logarithms
of this scale. The low $x$ OPE developed in~\cite{Balitsky:1995ub, Balitsky:1998kc, Balitsky:1998ya},
which we refer to as the shockwave framework, has a very different
form: 
\begin{equation}
J(z)J(0)\rightarrow C_{0}\left(z,Y_{c}\right)\mathcal{O}_{0}\left(Y_{c}\right)+\alpha_{s}C_{1}\left(z,Y_{c}\right)\mathcal{O}_{1}\left(Y_{c}\right)+...,\label{eq:LowxOPE}
\end{equation}
where $Y_{c}$ is a rapidity separation scale and where the spurious
rapidity divergence from the 1-loop correction to the $C_{n+1}$ coefficient
is cancelled by the Leading Logarithmic (LL) $Y_{c}$ evolution of
the $n$-th operator $\mathcal{O}_{n}$, and so on and so forth. Such
an evolution in $Y_{c}$ allows to resum large logarithms of $s$. 

It is possible to take the low $x$ limit in Eq.~(\ref{eq:colOPE})
and to match it to the first powers of the partonic hard scale in
Eq.~(\ref{eq:LowxOPE}). However, Eq.~(\ref{eq:colOPE}) is usually
valid for the first few powers at best while Eq.~(\ref{eq:LowxOPE})
is valid for all powers of $Q$. On the other side, so far only the
first few subleading $s^{-1/2}$ corrections to Eq.~(\ref{eq:LowxOPE})
have been computed~\cite{Altinoluk:2014oxa, Altinoluk:2015gia, Altinoluk:2015xuy, Agostini:2019avp, Balitsky:2015qba, Balitsky:2016dgz, Balitsky:2017flc, Balitsky:2017gis, Chirilli:2018kkw, Kovchegov:2015pbl, Kovchegov:2016zex} while Eq.~(\ref{eq:colOPE})
is valid for all powers of $s$.

Finding a general continuity between the moderate and low $x$ factorization schemes is very challenging due to the fact that the operators involved in these two schemes have different nature. While the moderate $x$ factorization
schemes involve operators which consist of parton fields with the appropriate gauge links, the low $x$ schemes involve full Wilson line operators. The main focus of this paper is to address this challenge.


For several processes and in several kinematic regimes, a matching
between the low $x$ Wilson line operators and some standard moderate
$x$ distributions have been found. Transverse Momentum Dependent
(TMD) distributions were first recovered in~\cite{Dominguez:2010xd, Dominguez:2011wm} 
via a so-called \textit{correlation} expansion, which was
extended in~\cite{Altinoluk:2018uax, Altinoluk:2018byz} for 3-particle final states and to infinite kinematic twist accuracy in~\cite{Altinoluk:2019fui}. The correspondance between low $x$ and TMD observables is the subject of many recent studies~\cite{Metz:2011wb, Akcakaya:2012si, Dumitru:2016jku, Marquet:2016cgx, Boer:2017xpy, Marquet:2017xwy}. For a review on TMD gluon distributions at small $x$, the reader is referred to~\cite{Petreska:2018cbf}. The off-forward generalization of TMD distributions (Generalized
TMD distributions, GTMD) and their Fourier transforms, the Wigner distributions,
were also found in~\cite{Hatta:2016dxp, Boussarie:2018zwg}. Generalized Parton Distributions (GPD), the off-forward extension of PDFs, were extracted via a twist
expansion in~\cite{Hatta:2017cte}. In this paper, we develop
a method to rewrite low $x$ physics in terms of TMD and Wigner distributions,
which are known to span PDFs and GPDs as well~\cite{Belitsky:2003nz, Lorce:2011kd}, allowing for a completely systematic rewriting of low $x$ observables in terms the distributions involved in the moderate
$x$ regime as well. Our formulation of low $x$ physics is that of
an infinite twist (G)TMD framework.

The question of gluon saturation is one of the most exciting topics
in low $x$ physics. While the original Balitsky-Fadin-Kuraev-Lipatov
(BFKL) description~\cite{Kuraev:1977fs, Balitsky:1978ic} did not involve saturation effects, the more recent
dipole~\cite{Mueller:1989st, Mueller:1993rr, Mueller:1994gb} and shockwave~\cite{Balitsky:1995ub, Balitsky:1998kc, Balitsky:1998ya} frameworks contain non-linear effects embedded in their evolution equation. More
strikingly, the effective Feynman rules and Wilson line operators
involved in the shockwave framework were shown to be perfectly compatible
with earlier results~\cite{McLerran:1993ni, McLerran:1993ka, McLerran:1994vd} describing scattering off
a heavy ion with large gluon occupancy taken into account as the very
starting point. The low $x$ description of scattering off nuclear
targets, known as the Color Glass Condensate (CGC)~\cite{Gelis:2010nm},
also has the exact same hierarchy of evolution equations as in~\cite{Balitsky:1995ub, Balitsky:1998kc, Balitsky:1998ya}. This evolution equation is called the Balitsky-Jalilian-Marian-Iancu-McLerran-Weigert-Leonidov-Kovner
(B-JIMWLK) evolution equation~\cite{JalilianMarian:1997jx, JalilianMarian:1997gr, JalilianMarian:1997dw, Kovner:1999bj, Kovner:2000pt, Weigert:2000gi, Iancu:2000hn, Ferreiro:2001qy} and in the mean field approximation it reduces 
to the Balitsky-Kovchegov (BK) equation \cite{Kovchegov:1999yj} which is known as the evolution
equation in the dipole framework.

Saturation is usually understood as a combination of two distinct effects: gluon recombinations
via non-linearities in the evolution equation due to $x$ being small~\cite{Gribov:1984tu},
and the importance of multiple scatterings due to the large gluon
occupancy number for dense targets~\cite{McLerran:1993ka, McLerran:1993ni, McLerran:1994vd}. 
In this paper, we give a new point of view on saturation in terms of TMD physics. In particular, we distinguish 3 origins of saturation, whose effects can be studied separately. Moreover, we discuss the linear BFKL limit as
a kinematic limit rather than a dilute limit.

We restrict ourselves to dilute-dense
collisions where either the projectile is a photon (with or without
virtuality), or where in a hadron-hadron collision the observed particles are forward enough for the
so-called hybrid factorization ansatz~\cite{Dumitru:2005gt, Altinoluk:2011qy} to apply. This
ansatz relies on the assumption that the incoming projectile parton
is produced close enough to the projectile hadron beam to be reliably
described via colinear factorization. Then the observables are described
as the convolution of a colinear parton distribution with a low $x$
amplitude where the incoming parton is treated as the projectile. The hybrid ansatz has been successfully applied to one-loop order~\cite{Chirilli:2011km, Chirilli:2012jd, Stasto:2013cha, Stasto:2014sea, Altinoluk:2014eka, Watanabe:2015tja, Ducloue:2016shw, Iancu:2016vyg, Ducloue:2017dit, Ducloue:2017mpb} 
For central production, a more involved formalism similar to the one developed in~\cite{Balitsky:2017flc, Balitsky:2017gis} should be used, but this is beyond the scope of this paper and we leave this case for future studies.

The paper is organized as follows. In Section~\ref{sec:Amplitudes},
we give the computation steps to rewrite low $x$ amplitudes in a
form that is compatible with an infinite twist TMD amplitude, involving
all kinematic twist corrections to the 1-body and 2-body (half-)operators.
In Section~\ref{sec:InclusiveXS}, we derive the cross section in
the inclusive case, involving 2-body, 3-body and 4-body TMD distributions.
We then show how PDFs appear in more inclusive observables and briefly 
discuss inclusive diffraction. In Section~\ref{sec:ExclusiveXS},
we derive the cross section in the exclusive case, which involves
a GTMD and show how GPDs appear for less exclusive observables.
In Section~\ref{sec:BFKL}, we show how the BFKL limit can be understood
as a kinematic limit in the Wandzura-Wilczek approximation, and we
give predictions beyond the WW approximation in terms of 2-body, 3-body
and 4-body unintegrated PDFs. Finally, in Section~\ref{sec:Saturation},
we discuss how saturation can be understood in terms of TMD physics
and how one kind of saturation could also appear in the kinematic
BFKL limit.

\section*{Notations and conventions}

We consider the most generic small $x$ limit: $s$ is assumed to
be much larger than any other scale, and our processes are assumed
to have at least one partonic hard scale $Q$. Any number of hard
or semi-hard scales can be involved.

We define lightcone directions $+$ and $-$ such that the projectile
(resp. target) has a large momentum along the $+$ direction $p_{0}^{+}\sim\sqrt{s}$
(resp. along the $-$ direction $P^{-}\sim\sqrt{s}$). We denote transverse
components as with a $\perp$ subscript in Minkowski space and by
bold characters in Euclidean space. Thus, the scalar product is written 
as
\begin{equation}
x\cdot y=x^{+}y^{-}+x^{-}y^{+}+x_{\perp}\cdot y_{\perp}=x^{+}y^{-}+x^{-}y^{+}-\boldsymbol{x}\cdot\boldsymbol{y}.\label{eq:SP}
\end{equation}
Our treatement of low $x$ physics relies on the covariant shockwave
effective approach which is very similar to the CGC approach.
Both of these frameworks are based on the separation of gluon fields in rapidity
space: the QCD Lagrangian is separated\footnote{For central production in hadron-hadron collisions, additional separations
would be introduced.} into \textit{fast} fields ($|k^{+}|>e^{-Y_{c}}p_{0}^{+}$) and \textit{slow}
fields ($|k^{+}|<e^{-Y_{c}}p_{0}^{+}$). We use the lightcone gauge
$A^{+}=0,$ in which slow fields have the form
\begin{equation}
A_{Y_{c}}^{\mu}(x) = \delta(x^{+})A_{Y_{c}}^{-}(x_{\perp})\delta^{\mu-}+O\left(m_{T}/\sqrt{s}\right),\label{eq:shockwave}
\end{equation}
where $m_{T}$ is a typical mass in the target. Note that the $\delta(x^+)$ function accounts for the fact that the field is peaked around $x^+=0$, and should be treated as such rather than an actual $\delta$-distribution.   The slow field is
then treated as an external field, which allows for multiple interactions
to be resummed into path-ordered Wilson lines. For a color representation
$R$, the finite Wilson line operators are defined as
\begin{equation}
\left[a^{+},b^{+}\right]_{\boldsymbol{x},Y_{c}}^{R}\equiv\mathcal{P}e^{ig\int_{a^{+}}^{b^{+}}dz^{+}T_{R}^{a}A_{Y_{c}}^{a-}\left(z^{+},\boldsymbol{x}\right)},\label{eq:WLdef}
\end{equation}
and the more standard infinite lines are defined as
\begin{equation}
U_{\boldsymbol{x},Y_{c}}^{R}=\left[-\infty,+\infty\right]_{\boldsymbol{x},Y_{c}}^{R}.\label{eq:Udef}
\end{equation}
Large logarithms of $s$ are resummed via the $Y_{c}$ evolution of
the Wilson line operators, given by the B-JIMWLK hierarchy of evolution
equations~\cite{Balitsky:1995ub, Balitsky:1998kc, Balitsky:1998ya, JalilianMarian:1997jx, JalilianMarian:1997gr, JalilianMarian:1997dw, Kovner:1999bj, Kovner:2000pt, Weigert:2000gi, Iancu:2000hn, Ferreiro:2001qy} , which can be written in a compact form
as the action of the JIMWLK Hamiltonian $H$ on a functional of Wilson
line operators
\begin{equation}
\frac{d}{dY_{c}}\left(U_{\boldsymbol{x}_{1},Y_{c}}^{R_{1}}...U_{\boldsymbol{x}_{n},Y_{c}}^{R_{n}}\right)=-H \cdot\left(U_{\boldsymbol{x}_{1},Y_{c}}^{R_{1}}...U_{\boldsymbol{x}_{n},Y_{c}}^{R_{n}}\right).\label{eq:JIMWLK}
\end{equation}
Then the observables are given as the convolution of the hard part
$\mathcal{H}_{Y_{c}}\left(\boldsymbol{x}_{1},...,\boldsymbol{x}_{n}\right)$,
obtained through effective Feynman rules in the external slow (classical)
field with a lower rapidity cutoff $Y_{c}$ for the fast (quantum)
gluon fields, and the action of Wilson line operators at rapidity
$Y_{c}$ on target states $\left\langle P^{\left(\prime\right)}|P\right\rangle $:

\begin{equation}
\mathcal{A}=\int d^{2}\boldsymbol{x}_{1}...d^{2}\boldsymbol{x}_{n}\mathcal{H}_{Y_{c}}\left(\boldsymbol{x}_{1},...,\boldsymbol{x}_{n}\right)\frac{\left\langle P^{\left(\prime\right)}\left|U_{\boldsymbol{x}_{1},Y_{c}}^{R_{1}}...U_{\boldsymbol{x}_{n},Y_{c}}^{R_{n}}\right|P\right\rangle }{\left\langle P|P\right\rangle },\label{eq:convolution}
\end{equation}
where for exclusive observables $\mathcal{A}$ is the amplitude and
the matrix element is off-diagonal, and for inclusive observables
$\mathcal{A}$ is the cross section and the matrix element is diagonal.
Hereafter, we drop the subscript $Y_{c}$ subscripts for convenience. We normalize our target states such that
\begin{equation}
\left\langle P^{\prime}|P\right\rangle =2P^{-}\left(2\pi\right)^{3}\delta\left(P^{\prime-}-P^{-}\right)\delta^{2}\left(\boldsymbol{P}^{\prime}-\boldsymbol{P}\right),\label{eq:TargetNorm}
\end{equation}
and use extensively relations similar to

\begin{align}
 & \int dx_{1}^{+}dx_{2}^{+}d^{2}\boldsymbol{x}_{1}d^{2}\boldsymbol{x}_{2}\left.\frac{\left\langle P\left|\left[x_{2}^{+},x_{1}^{+}\right]_{\boldsymbol{x}_{2}}F^{i-}\left(x_{1}\right)\left[x_{1}^{+},x_{2}^{+}\right]_{\boldsymbol{x}_{1}}F^{j-}\left(x_{2}\right)\right|P\right\rangle }{\left\langle P|P\right\rangle }\right|_{x_{1,2}^{-}=0}\nonumber \\
 & =\frac{1}{2P^{-}}\int dr^{+}d^{2}\boldsymbol{r}\left.\left\langle P\left|\left[0^{+},r^{+}\right]_{\boldsymbol{0}}F^{i-}\left(r\right)\left[r^{+},0^{+}\right]_{\boldsymbol{r}}F^{j-}\left(0\right)\right|P\right\rangle \right|_{r^{-}=0},\label{eq:TransInv-1}
\end{align}
as a result of normalization (\ref{eq:TargetNorm}) and translation invariance. We also use the fact that in lightcone gauge $A^+=0$ and in the eikonal approximation given in Eq.~(\ref{eq:shockwave}), the \textit{slow} gluon field $F^{i-}$ simply reads
\begin{equation}
F^{i-}(x) = \partial^i A^-(x).\label{Fij}
\end{equation}
Finally, the connection to standard parton distributions is always
obtained by using small $x$ limits of these distributions as given
by relations of type:
\begin{align}
 & \int\! dr^{+}d^{2}\boldsymbol{r}e^{ixP^{-}r^{+}}\!\left.\left\langle P^{\prime}\left|F^{i-}\left(r\right)\mathcal{U}_{\left[r,0\right]}^{\pm}F^{j-}\left(0\right)\mathcal{U}_{\left[0,r\right]}^{\pm}\right|P\right\rangle \right|_{r^{-}=0}\label{eq:xeq0} \\
 & \rightarrow\!\int\! dr^{+}d^{2}\boldsymbol{r}\!\left.\left\langle P^{\prime}\left|F^{i-}\left(r\right)\left[r^{+},\pm\infty\right]_{\boldsymbol{r}}\left[\pm\infty,0^{+}\right]_{\boldsymbol{0}}F^{j-}\left(0\right)\left[0^{+},\pm\infty\right]_{\boldsymbol{0}}\left[\pm\infty,r^{+}\right]_{\boldsymbol{r}}\right|P\right\rangle \right|_{r^{-}=0},\nonumber
\end{align}
where $\mathcal{U}_{\left[x,y\right]}^{\pm}$ are staple gauge links:
\begin{equation}
\mathcal{U}_{\left[x,y\right]}^{\pm}=\left[\left(x^{+},x^{-}\!,\boldsymbol{x}\right),\left(\pm\infty,x^{-}\!,\boldsymbol{x}\right)\right]\left[\left(\pm\infty,x^{-}\!,\boldsymbol{x}\right),\left(\pm\infty,x^{-}\!,\boldsymbol{y}\right)\right]\left[\left(\pm\infty,x^{-}\!,\boldsymbol{y}\right),\left(y^{+},x^{-},\boldsymbol{y}\right)\right],\label{eq:staple}
\end{equation}
whose transverse parts are subeikonal in $A^{+}=0$ gauge.

\section{Low $x$ amplitudes as (G)TMD amplitudes\label{sec:Amplitudes}}

All the computation steps which we perform here would apply
for generic shockwave amplitudes. We restrict ourselves to processes
with 1 incoming particle of momentum $p_0$ in color representation $R_0$ and 2 outgoing particles of respective momenta $p_1$ and $p_2$ and in respective color representations $R_1$ and $R_2$, in the external field of
a hadronic target. In the shockwave and CGC formulations of low $x$ physics,
the amplitude for such a process has the form~\cite{Altinoluk:2019fui}
\begin{align}
\mathcal{A} & =\left(2\pi\right)\delta\left(p_{1}^{+}+p_{2}^{+}-p_{0}^{+}\right)\int\!d^{2}\boldsymbol{b}\,d^{2}\boldsymbol{r}\,e^{-i\left(\boldsymbol{q}\cdot\boldsymbol{r}\right)-i\left(\boldsymbol{k}\cdot\boldsymbol{b}\right)}\mathcal{H}\left(\boldsymbol{r}\right)\label{eq:GenAmp}\\
 & \times\left[\left(U_{\boldsymbol{b}+\bar{z}\boldsymbol{r}}^{R_{1}}T^{R_{0}}U_{\boldsymbol{b}-z\boldsymbol{r}}^{R_{2}}\right)-\left(U_{\boldsymbol{b}}^{R_{1}}T^{R_{0}}U_{\boldsymbol{b}}^{R_{2}}\right)\right],\nonumber 
\end{align}
where $\mathcal{H}\left(\boldsymbol{r}\right)$ is the hard part, and we defined 
\begin{equation} \boldsymbol{k}\equiv \boldsymbol{p}_1+\boldsymbol{p}_2, \label{kdef} 
\end{equation}
and
\begin{equation} \boldsymbol{q}\equiv \frac{p_1^+\boldsymbol{p}_1-p_2^+\boldsymbol{p}_2}{p_1^+ +p_2^+}, \label{qdef}
\end{equation}
and where
\begin{equation}
z\equiv\frac{p_1+}{p_0^+},\quad \bar{z}\equiv\frac{p_2^+}{p_0^+}=1-z. \label{zdef}
\end{equation}
The study we perform here is independent of $\mathcal{H}$ and thus
valid for any process with 2 outgoing particles, with or without masses
or virtualities. Note that final state partons can hadronize for example
via fragmentation functions, distribution amplitudes or NRQCD, without changing the validity of present study either. Also note that the
true amplitude is given by the action of the Wilson line operators
on target states, which we will introduce later in Section~\ref{sec:InclusiveXS}
in the inclusive case and in Section~\ref{sec:ExclusiveXS} in the
exclusive case. For the moment, it is enough to keep the amplitude
as an operator.

The equivalence we want to prove here relies on the following rewriting
of Wilson lines in terms of their derivatives:
\begin{align}
U_{\boldsymbol{b}+\bar{z}\boldsymbol{r}}^{R_{1}} & =U_{\boldsymbol{b}}^{R_{1}}-i r_\perp^{\alpha}\!\int\!\frac{d^{2}\boldsymbol{k}_{1}}{\left(2\pi\right)^{2}}\!\int\!d^{2}\boldsymbol{b}_{1}\,e^{-i\boldsymbol{k}_{1}\cdot\left(\boldsymbol{b}_{1}-\boldsymbol{b}\right)}\frac{e^{i\bar{z}\left(\boldsymbol{k}_{1}\cdot\boldsymbol{r}\right)}-1}{\left(\boldsymbol{k}_{1}\cdot\boldsymbol{r}\right)}\left(\partial_{\alpha}U_{\boldsymbol{b}_{1}}^{R_{1}}\right),\label{eq:TrickR1}
\end{align}
and 
\begin{align}
U_{\boldsymbol{b}-z\boldsymbol{r}}^{R_{2}} & =U_{\boldsymbol{b}}^{R_{2}}-ir_\perp^\alpha\!\int\!\frac{d^{2}\boldsymbol{k}_{2}}{\left(2\pi\right)^{2}}\!\int\!d^{2}\boldsymbol{b}_{2}\,e^{-i\boldsymbol{k}_{2}\cdot\left(\boldsymbol{b}_{2}-\boldsymbol{b}\right)}\frac{e^{-iz\left(\boldsymbol{k}_{2}\cdot\boldsymbol{r}\right)}-1}{\left(\boldsymbol{k}_{2}\cdot\boldsymbol{r}\right)}\left(\partial_{\alpha}U_{\boldsymbol{b}_{2}}^{R_{2}}\right).\label{eq:TrickR2}
\end{align}
Indeed, derivatives of Wilson line operators are the main quantities to
consider when trying to match the TMD formalism, as shown in~\cite{Dominguez:2011wm}. The derivative of a fundamental line is given by
\begin{equation}
\left(\partial_{i}U_{\boldsymbol{b}_{1}}^{R}\right)=ig_{s}T_{R}^{a}\int_{-\infty}^{+\infty}db_{1}^{+}\left[-\infty,b_{1}^{+}\right]_{\boldsymbol{b}_{1}}^{R}F_{a}^{i-}\left(b_{1}^{+},0^{-},\boldsymbol{b}_{1}\right)\left[b_{1}^{+},+\infty\right]_{\boldsymbol{b}_{1}}^{R},\label{eq:DiU}
\end{equation}
which allows to identify the $F_{a}^{i-}\left(b_{1}\right)$ field
as the actual gluon field in a TMD operator. At moderate $x$ this
gluon would be isolated from the slow gluons in the gauge links. The
amplitude in Eq. (\ref{eq:GenAmp}) can be rewritten as a sum of three pieces:
\begin{align}
\mathcal{A} & \equiv\mathcal{A}_{g}+\mathcal{A}_{k}^{\left(1\right)}+\mathcal{A}_{k}^{\left(2\right)},\label{eq:ExpandedAmp-1}
\end{align}
where the three pieces are defined respectively as
\begin{align}
\mathcal{A}_{g} & =\left(2\pi\right)\delta\left(p_{1}^{+}+p_{2}^{+}-p_{0}^{+}\right)\int\!d^{2}\boldsymbol{b}\,d^{2}\boldsymbol{r}\,e^{-i\left(\boldsymbol{q}\cdot\boldsymbol{r}\right)-i\left(\boldsymbol{k}\cdot\boldsymbol{b}\right)}\mathcal{H}\left(\boldsymbol{r}\right)\label{eq:ExpandedAmp}\\
 & \times\left(U_{\boldsymbol{b}+\bar{z}\boldsymbol{r}}^{R_{1}}-U_{\boldsymbol{b}}^{R_{1}}\right)T^{R_{0}}\left(U_{\boldsymbol{b}-z\boldsymbol{r}}^{R_{2}}-U_{\boldsymbol{b}}^{R_{2}}\right),\nonumber 
\end{align}
\begin{align}
\mathcal{A}_{k}^{\left(1\right)} & =\left(2\pi\right)\delta\left(p_{1}^{+}+p_{2}^{+}-p_{0}^{+}\right)\int\!d^{2}\boldsymbol{b}\,d^{2}\boldsymbol{r}\,e^{-i\left(\boldsymbol{q}\cdot\boldsymbol{r}\right)-i\left(\boldsymbol{k}\cdot\boldsymbol{b}\right)}\mathcal{H}\left(\boldsymbol{r}\right)\label{eq:ExpandedAmp-2}\\
 & \times\left(U_{\boldsymbol{b}+\bar{z}\boldsymbol{r}}^{R_{1}}-U_{\boldsymbol{b}}^{R_{1}}\right)T^{R_{0}}U_{\boldsymbol{b}}^{R_{2}},\nonumber 
\end{align}
and
\begin{align}
\mathcal{A}_{k}^{\left(2\right)} & =\left(2\pi\right)\delta\left(p_{1}^{+}+p_{2}^{+}-p_{0}^{+}\right)\int\!d^{2}\boldsymbol{b}\,d^{2}\boldsymbol{r}\,e^{-i\left(\boldsymbol{q}\cdot\boldsymbol{r}\right)-i\left(\boldsymbol{k}\cdot\boldsymbol{b}\right)}\mathcal{H}\left(\boldsymbol{r}\right)\label{eq:ExpandedAmp-3}\\
 & \times U_{\boldsymbol{b}}^{R_{1}}T^{R_{0}}\left(U_{\boldsymbol{b}-z\boldsymbol{r}}^{R_{2}}-U_{\boldsymbol{b}}^{R_{2}}\right).\nonumber 
\end{align}
The first piece is easy to rewrite thanks to Eqs.~(\ref{eq:TrickR1})
and (\ref{eq:TrickR2}):
\begin{align}
\mathcal{A}_{g} & =\left(2\pi\right)\delta\left(p_{1}^{+}+p_{2}^{+}-p_{0}^{+}\right)\int\!\frac{d^{2}\boldsymbol{k}_{1}}{\left(2\pi\right)^{2}}\frac{d^{2}\boldsymbol{k}_{2}}{\left(2\pi\right)^{2}}\left(2\pi\right)^{2}\delta^{2}\left(\boldsymbol{k}_{1}+\boldsymbol{k}_{2}-\boldsymbol{k}\right)\nonumber \\
 & \times\int\!d^{2}\boldsymbol{b}_{1}d^{2}\boldsymbol{b}_{2}\,e^{-i\left(\boldsymbol{k}_{1}\cdot\boldsymbol{b}_{1}\right)-i\left(\boldsymbol{k}_{2}\cdot\boldsymbol{b}_{2}\right)}\left(\partial_{i}U_{\boldsymbol{b}_{1}}^{R_{1}}\right)T^{R_{0}}\left(\partial_{j}U_{\boldsymbol{b}_{2}}^{R_{2}}\right)\label{eq:AGfin}\\
 & \times\int\!d^{2}\boldsymbol{r}\,e^{-i\left(\boldsymbol{q}\cdot\boldsymbol{r}\right)}\left[-\boldsymbol{r}^{i}\boldsymbol{r}^{j}\mathcal{H}\left(\boldsymbol{r}\right)\frac{\left(e^{i\bar{z}\left(\boldsymbol{k}_{1}\cdot\boldsymbol{r}\right)}-1\right)\left(e^{-iz\left(\boldsymbol{k}_{2}\cdot\boldsymbol{r}\right)}-1\right)}{\left(\boldsymbol{k}_{1}\cdot\boldsymbol{r}\right)\left(\boldsymbol{k}_{2}\cdot\boldsymbol{r}\right)}\right].\nonumber 
\end{align}
This contribution is perfectly compatible with an all-kinematic-twists-resummed
TMD amplitude for the first subleading-twist TMD half-operator.

The second and third pieces $\mathcal{A}_{k}^{\left(1\right)}$ and
$\mathcal{A}_{k}^{\left(2\right)}$ contain both 1-gluon and 2-gluon
contributions. The 1-gluon contributions were extracted and resummed
in~\cite{Altinoluk:2019fui}, then compared to predictions from the kinematic-twist-resummed
TMD framework developped in~\cite{Kotko:2015ura, vanHameren:2016ftb} for several concrete
examples. Let us recall the method which was used, and resum the 2-gluon
contributions as well. The Taylor expanded form of the
Wilson line operator involved in $\mathcal{A}_{k}^{\left(1\right)}$ can be written as
\begin{equation}
\left(U_{\boldsymbol{b}+\bar{z}\boldsymbol{r}}^{R_{1}}-U_{\boldsymbol{b}}^{R_{1}}\right)T^{R_{0}}U_{\boldsymbol{b}}^{R_{2}}=\sum_{n=1}^{\infty}\frac{\bar{z}^{n}}{n!}\left[\left(r_{\perp}\cdot\partial_{\perp}\right)^{n}U_{\boldsymbol{b}}^{R_{1}}\right]T^{R_{0}}U_{\boldsymbol{b}}^{R_{2}}\equiv\sum_{n=1}^{\infty}\mathcal{U}_{n}.\label{eq:Taylor1}
\end{equation}
Keeping in mind that $\mathcal{U}_{n}$ will be integrated as $\int\!d^{2}\boldsymbol{b}\,e^{-i\left(\boldsymbol{k}\cdot\boldsymbol{b}\right)}\,\mathcal{U}_{n}$,
we can use integrations by parts to write
\begin{equation}
\mathcal{U}_{n}=\frac{-i\bar{z}\left(k_{\perp}\cdot r_{\perp}\right)}{n}\mathcal{U}_{n-1}-\frac{\bar{z}^{n}}{n!}\left[\left(r_{\perp}\cdot\partial_{\perp}\right)^{n-1}U_{\boldsymbol{b}}^{R_{1}}\right]T^{R_{0}}\left(r_{\perp}\cdot\partial_{\perp}\right)U_{\boldsymbol{b}}^{R_{2}}.\label{eq:IbP1}
\end{equation}
Then an easy recursion shows
\begin{align}
\mathcal{U}_{n} & =\frac{\left[-i\bar{z}\left(k_{\perp}\cdot r_{\perp}\right)\right]^{n-1}}{n!}\mathcal{U}_{1}\label{eq:UnRec}\\
 & -\boldsymbol{r}^{i}\boldsymbol{r}^{j}\frac{\bar{z}^{n}}{n!}\sum_{m=1}^{n-1}\left[i\left(\boldsymbol{k}\cdot\boldsymbol{r}\right)\right]^{n-1-m}\left[\left(r_{\perp}\cdot\partial_{\perp}\right)^{m-1}\left(\partial^{i}U_{\boldsymbol{b}}^{R_{1}}\right)\right]T^{R_{0}}\left(\partial^{j}U_{\boldsymbol{b}}^{R_{2}}\right).\nonumber 
\end{align}
Using
\begin{equation}
\left(r_{\perp}\cdot\partial_{\perp}\right)^{m-1}\left(\partial_{i}U_{\boldsymbol{b}}^{R_{1}}\right)=\int\!d^{2}\boldsymbol{b}_{1}\int\!\frac{d^{2}\boldsymbol{k}_{1}}{\left(2\pi\right)^{2}}\left(i\boldsymbol{k}_{1}\cdot\boldsymbol{r}\right)^{m-1}e^{i\boldsymbol{k}_{1}\cdot\left(\boldsymbol{b}-\boldsymbol{b}_{1}\right)}\left(\partial_{i}U_{\boldsymbol{b}_{1}}^{R_{1}}\right),\label{eq:DmU1}
\end{equation}
and 
\begin{equation}
\left(\partial_{j}U_{\boldsymbol{b}}^{R_{2}}\right)=\int\!d^{2}\boldsymbol{b}_{2}\int\!\frac{d^{2}\boldsymbol{k}_{2}}{\left(2\pi\right)^{2}}e^{i\boldsymbol{k}_{2}\cdot\left(\boldsymbol{b}-\boldsymbol{b}_{2}\right)}\left(\partial_{j}U_{\boldsymbol{b}_{2}}^{R_{2}}\right),\label{eq:U2}
\end{equation}
one can obtain
\begin{align}
\mathcal{U}_{n} & =\frac{\left[i\bar{z}\left(\boldsymbol{k}\cdot\boldsymbol{r}\right)\right]^{n-1}}{n!}\mathcal{U}_{1}-\boldsymbol{r}^{i}\boldsymbol{r}^{j}\frac{\bar{z}^{2}}{n!}\int\!\frac{d^{2}\boldsymbol{k}_{1}}{\left(2\pi\right)^{2}}\frac{d^{2}\boldsymbol{k}_{2}}{\left(2\pi\right)^{2}}e^{i\left(\boldsymbol{k}_{1}+\boldsymbol{k}_{2}\right)\cdot\boldsymbol{b}}\int\!d^{2}\boldsymbol{b}_{1}d^{2}\boldsymbol{b}_{2}\,e^{-i\left(\boldsymbol{k}_{1}\cdot\boldsymbol{b}_{1}\right)-i\left(\boldsymbol{k}_{2}\cdot\boldsymbol{b}_{2}\right)}\nonumber \\
 & \times\sum_{m=1}^{n-1}\left[i\bar{z}\left(\boldsymbol{k}\cdot\boldsymbol{r}\right)\right]^{n-1-m}\left(i\bar{z}\boldsymbol{k}_{1}\cdot\boldsymbol{r}\right)^{m-1}\left(\partial_{i}U_{\boldsymbol{b}_{1}}^{R_{1}}\right)T^{R_{0}}\left(\partial_{j}U_{\boldsymbol{b}_{2}}^{R_{2}}\right).\label{eq:UnRes}
\end{align}
A final resummation, using the relations
\begin{equation}
\sum_{n=1}^{\infty}\frac{X^{n-1}}{n!}=\frac{e^{X}-1}{X},\label{eq:Sum1}
\end{equation}
and

\begin{equation}
\sum_{n=1}^{\infty}\sum_{m=1}^{n-1}\frac{X^{m-1}Y^{n-1-m}}{n!}=\frac{Y\left(e^{X}-1\right)-X\left(e^{Y}-1\right)}{XY\left(X-Y\right)},\label{eq:Sum2}
\end{equation}
leads to
\begin{align}
 & \left(U_{\boldsymbol{b}+\bar{z}\boldsymbol{r}}^{R_{1}}-U_{\boldsymbol{b}}^{R_{1}}\right)T^{R_{0}}U_{\boldsymbol{b}}^{R_{2}}=i\boldsymbol{r}^{i}\frac{e^{i\bar{z}\left(\boldsymbol{k}\cdot\boldsymbol{r}\right)}-1}{\left(\boldsymbol{k}\cdot\boldsymbol{r}\right)}\left(\partial^{i}U_{\boldsymbol{b}}^{R_{1}}\right)T^{R_{0}}U_{\boldsymbol{b}}^{R_{2}}\label{eq:U1fin}\\
 & -\boldsymbol{r}^{i}\boldsymbol{r}^{j}\!\int\!\frac{d^{2}\boldsymbol{k}_{1}}{\left(2\pi\right)^{2}}\frac{d^{2}\boldsymbol{k}_{2}}{\left(2\pi\right)^{2}}e^{i\left(\boldsymbol{k}_{1}+\boldsymbol{k}_{2}\right)\cdot\boldsymbol{b}}\int\!d^{2}\boldsymbol{b}_{1}d^{2}\boldsymbol{b}_{2}\,e^{-i\left(\boldsymbol{k}_{1}\cdot\boldsymbol{b}_{1}\right)-i\left(\boldsymbol{k}_{2}\cdot\boldsymbol{b}_{2}\right)}\nonumber \\
 & \times\frac{\left(\boldsymbol{k}\cdot\boldsymbol{r}\right)\left(e^{i\bar{z}\left(\boldsymbol{k}_{1}\cdot\boldsymbol{r}\right)}-1\right)-\left(\boldsymbol{k}_{1}\cdot\boldsymbol{r}\right)\left(e^{i\bar{z}\left(\boldsymbol{k}\cdot\boldsymbol{r}\right)}-1\right)}{\left(\boldsymbol{k}_{1}\cdot\boldsymbol{r}\right)\left(\boldsymbol{k}\cdot\boldsymbol{r}\right)\left(\boldsymbol{k}-\boldsymbol{k}_{1}\right)\cdot\boldsymbol{r}}\left(\partial_{i}U_{\boldsymbol{b}_{1}}^{R_{1}}\right)T^{R_{0}}\left(\partial_{j}U_{\boldsymbol{b}_{2}}^{R_{2}}\right).\nonumber 
\end{align}
Plugging Eq.(\ref{eq:U1fin}) into Eq.(\ref{eq:ExpandedAmp}) finally yields
\begin{align}
\mathcal{A}_{k}^{\left(1\right)} & =\left(2\pi\right)\delta\left(p_{1}^{+}+p_{2}^{+}-p_{0}^{+}\right)\!\int\!d^{2}\boldsymbol{b}\,e^{-i\left(\boldsymbol{k}\cdot\boldsymbol{b}\right)}\left(\partial^{i}U_{\boldsymbol{b}}^{R_{1}}\right)T^{R_{0}}U_{\boldsymbol{b}}^{R_{2}}\nonumber \\
 & \times\left[i\!\int\!d^{2}\boldsymbol{r}\,e^{-i\left(\boldsymbol{q}\cdot\boldsymbol{r}\right)}\boldsymbol{r}^{i}\mathcal{H}\left(\boldsymbol{r}\right)\left(\frac{e^{i\bar{z}\left(\boldsymbol{k}\cdot\boldsymbol{r}\right)}-1}{\left(\boldsymbol{k}\cdot\boldsymbol{r}\right)}\right)\right]\nonumber \\
 & +\left(2\pi\right)\delta\left(p_{1}^{+}+p_{2}^{+}-p_{0}^{+}\right)\int\!\frac{d^{2}\boldsymbol{k}_{1}}{\left(2\pi\right)^{2}}\frac{d^{2}\boldsymbol{k}_{2}}{\left(2\pi\right)^{2}}\left(2\pi\right)^{2}\delta\left(\boldsymbol{k}_{1}+\boldsymbol{k}_{2}-\boldsymbol{k}\right)\label{eq:AK1fin}\\
 & \times\int\!d^{2}\boldsymbol{b}_{1}d^{2}\boldsymbol{b}_{2}\,e^{-i\left(\boldsymbol{k}_{1}\cdot\boldsymbol{b}_{1}\right)-i\left(\boldsymbol{k}_{2}\cdot\boldsymbol{b}_{2}\right)}\left(\partial^{i}U_{\boldsymbol{b}_{1}}^{R_{1}}\right)T^{R_{0}}\left(\partial^{j}U_{\boldsymbol{b}_{2}}^{R_{2}}\right)\nonumber \\
 & \times\left[\int\!d^{2}\boldsymbol{r}\,e^{-i\left(\boldsymbol{q}\cdot\boldsymbol{r}\right)}\boldsymbol{r}^{i}\boldsymbol{r}^{j}\mathcal{H}\left(\boldsymbol{r}\right)\frac{\left(\boldsymbol{k}_{1}\cdot\boldsymbol{r}\right)\left(e^{i\bar{z}\left(\boldsymbol{k}\cdot\boldsymbol{r}\right)}-1\right)-\left(\boldsymbol{k}\cdot\boldsymbol{r}\right)\left(e^{i\bar{z}\left(\boldsymbol{k}_{1}\cdot\boldsymbol{r}\right)}-1\right)}{\left(\boldsymbol{k}_{1}\cdot\boldsymbol{r}\right)\left(\boldsymbol{k}_{2}\cdot\boldsymbol{r}\right)\left(\boldsymbol{k}\cdot\boldsymbol{r}\right)}\right],\nonumber 
\end{align}
where a 1-gluon contribution and a 2-gluon contribution were explicitely
extracted and power-resummed. Applying exactly the same method to
the remaining piece, we obtain
\begin{align}
\mathcal{A}_{k}^{\left(2\right)} & =\left(2\pi\right)\delta\left(p_{1}^{+}+p_{2}^{+}-p_{0}^{+}\right)\!\int\!d^{2}\boldsymbol{b}\,e^{-i\left(\boldsymbol{k}\cdot\boldsymbol{b}\right)}U_{\boldsymbol{b}}^{R_{1}}T^{R_{0}}\left(\partial^{i}U_{\boldsymbol{b}}^{R_{2}}\right)\nonumber \\
 & \times\left[i\!\int\!d^{2}\boldsymbol{r}\,e^{-i\left(\boldsymbol{q}\cdot\boldsymbol{r}\right)}\boldsymbol{r}^{i}\mathcal{H}\left(\boldsymbol{r}\right)\frac{e^{-iz\left(\boldsymbol{k}\cdot\boldsymbol{r}\right)}-1}{\left(\boldsymbol{k}\cdot\boldsymbol{r}\right)}\right]\nonumber \\
 & +\left(2\pi\right)\delta\left(p_{1}^{+}+p_{2}^{+}-p_{0}^{+}\right)\int\!\frac{d^{2}\boldsymbol{k}_{1}}{\left(2\pi\right)^{2}}\frac{d^{2}\boldsymbol{k}_{2}}{\left(2\pi\right)^{2}}\left(2\pi\right)^{2}\delta^{2}\left(\boldsymbol{k}_{1}+\boldsymbol{k}_{2}-\boldsymbol{k}\right)\label{eq:AK2fin}\\
 & \times\int\!d^{2}\boldsymbol{b}_{1}d^{2}\boldsymbol{b}_{2}\,e^{-i\left(\boldsymbol{k}_{1}\cdot\boldsymbol{b}_{1}\right)-i\left(\boldsymbol{k}_{2}\cdot\boldsymbol{b}_{2}\right)}\left(\partial^{i}U_{\boldsymbol{b}_{1}}^{R_{1}}\right)T^{R_{0}}\left(\partial^{j}U_{\boldsymbol{b}_{2}}^{R_{2}}\right)\nonumber \\
 & \times\left[\int\!d^{2}\boldsymbol{r}\,e^{-i\left(\boldsymbol{q}\cdot\boldsymbol{r}\right)}\mathcal{H}\left(\boldsymbol{r}\right)\boldsymbol{r}^{i}\boldsymbol{r}^{j}\frac{\left(\boldsymbol{k}_{2}\cdot\boldsymbol{r}\right)\left(e^{-iz\left(\boldsymbol{k}\cdot\boldsymbol{r}\right)}-1\right)-\left(\boldsymbol{k}\cdot\boldsymbol{r}\right)\left(e^{-iz\left(\boldsymbol{k}_{2}\cdot\boldsymbol{r}\right)}-1\right)}{\left(\boldsymbol{k}_{1}\cdot\boldsymbol{r}\right)\left(\boldsymbol{k}_{2}\cdot\boldsymbol{r}\right)\left(\boldsymbol{k}\cdot\boldsymbol{r}\right)}\right].\nonumber 
\end{align}
We can finally gather the 1-gluon and 2-gluon amplitudes. The 1-gluon amplitude reads:
\begin{align}
\mathcal{A}_{1} & =\left(2\pi\right)\delta\left(p_{1}^{+}+p_{2}^{+}-p_{0}^{+}\right)\!\int\!d^{2}\boldsymbol{b}\,e^{-i\left(\boldsymbol{k}\cdot\boldsymbol{b}\right)}(-i)\!\int\!d^{2}\boldsymbol{r}\,e^{-i\left(\boldsymbol{q}\cdot \boldsymbol{r}\right)}r_\perp^\alpha\mathcal{H}\left(\boldsymbol{r}\right)\label{eq:1Gfin}\\
 & \times\left[\left(\frac{e^{i\bar{z}\left(\boldsymbol{k}\cdot\boldsymbol{r}\right)}-1}{\left(\boldsymbol{k}\cdot\boldsymbol{r}\right)}\right)\left(\partial_\alpha U_{\boldsymbol{b}}^{R_{1}}\right)T^{R_{0}}U_{\boldsymbol{b}}^{R_{2}}+\left(\frac{e^{-iz\left(\boldsymbol{k}\cdot\boldsymbol{r}\right)}-1}{\left(\boldsymbol{k}\cdot\boldsymbol{r}\right)}\right)U_{\boldsymbol{b}}^{R_{1}}T^{R_{0}}\left(\partial_\alpha U_{\boldsymbol{b}}^{R_{2}}\right)\right].\nonumber 
\end{align}
Single-scattering contributions like those in Eq.~(\ref{eq:1Gfin}) were extracted for explicit processes
in~\citep{Altinoluk:2019fui} and the consistency of the results was checked
by comparing single-scattering cross sections derived with our method
to those obtained in the so-called improved TMD formalism, which is
a method to incorporate kinematic twists in TMD factorization. A perfect
match was found for all processes considered. \\
The 2-gluon amplitude is given by:
\begin{align}
\mathcal{A}_{2} & =\left(2\pi\right)\delta\left(p_{1}^{+}+p_{2}^{+}-p_{0}^{+}\right)\!\int\!\frac{d^{2}\boldsymbol{k}_{1}}{\left(2\pi\right)^{2}}\frac{d^{2}\boldsymbol{k}_{2}}{\left(2\pi\right)^{2}}\left(2\pi\right)^{2}\delta^{2}\left(\boldsymbol{k}_{1}+\boldsymbol{k}_{2}-\boldsymbol{k}\right)\nonumber \\
 & \times\int d^{2}\boldsymbol{b}_{1}d^{2}\boldsymbol{b}_{2}e^{-i\left(\boldsymbol{k}_{1}\cdot\boldsymbol{b}_{1}\right)-i\left(\boldsymbol{k}_{2}\cdot\boldsymbol{b}_{2}\right)}\left(\partial^{i}U_{\boldsymbol{b}_{1}}^{R_{1}}\right)T^{R_{0}}\left(\partial^{j}U_{\boldsymbol{b}_{2}}^{R_{2}}\right)\label{eq:2Gfin}\\
 & \times\left[-\int d^{2}\boldsymbol{r}e^{-i\left(\boldsymbol{q}\cdot\boldsymbol{r}\right)}\boldsymbol{r}^{i}\boldsymbol{r}^{j}\mathcal{H}\left(\boldsymbol{r}\right)\left(\frac{e^{-iz\left(\boldsymbol{k}\cdot\boldsymbol{r}\right)}}{\left(\boldsymbol{k}\cdot\boldsymbol{r}\right)}\frac{e^{i\left(\boldsymbol{k}_{1}\cdot\boldsymbol{r}\right)}-1}{\left(\boldsymbol{k}_{1}\cdot\boldsymbol{r}\right)}+\frac{e^{i\bar{z}\left(\boldsymbol{k}\cdot\boldsymbol{r}\right)}}{\left(\boldsymbol{k}\cdot\boldsymbol{r}\right)}\frac{e^{-i\left(\boldsymbol{k}_{2}\cdot\boldsymbol{r}\right)}-1}{\left(\boldsymbol{k}_{2}\cdot\boldsymbol{r}\right)}\right)\right].\nonumber 
\end{align}
The crucial point to note is that Eqs.~(\ref{eq:1Gfin}, \ref{eq:2Gfin})
sum up exactly to the low $x$ amplitude in Eq.~(\ref{eq:GenAmp}). As
a result, this shows that any low $x$ amplitude of the form of Eq.~(\ref{eq:GenAmp})
can be rewritten as the sum of all kinematic twist corrections to
the single-scattering TMD amplitude and to the double-scattering
(G)TMD amplitude (i.e. the first genuine twist correction). The notable
absence of triple or higher scattering amplitudes is due to the eikonal
approximation: a contribution with two derivatives hitting the same
line i.e. with 2 low $x$ TMD gluons hitting the same parton, constitutes
a gauge invariance fixing contribution: it was already taken into account either as part of a
gauge link or as a kinematic twist correction. We thus expect a low $x$ amplitude
with $n$ final state particles to have at most an $n$-scattering
operator in its amplitude in the eikonal approximation. With subeikonal
corrections, one could have higher genuine twist contributions.

In principle, Eqs.~(\ref{eq:1Gfin}, \ref{eq:2Gfin}) answer the
long-sought equivalence  between low $x$ and moderate $x$ formulations
of factorization: modern formulations of low $x$ amplitudes can be rewritten as the sum over all twists of (G)TMD amplitudes in their small $x$
limit.

For the sake of the compactness of the notations let us introduce
\begin{equation}
\mathcal{I}_{\mathcal{H}}^{i}\left(\boldsymbol{q},\boldsymbol{p}\right)\equiv i\!\int\!d^{2}\boldsymbol{r}\,e^{-i\left(\boldsymbol{q}\cdot\boldsymbol{r}\right)}\boldsymbol{r}^{i}\mathcal{H}\left(\boldsymbol{r}\right)\left(\frac{e^{i\left(\boldsymbol{p}\cdot\boldsymbol{r}\right)}-1}{\left(\boldsymbol{p}\cdot\boldsymbol{r}\right)}\right),\label{eq:Ints1}
\end{equation}
and
\begin{equation}
\mathcal{J}_{\mathcal{H}}^{ij}\left(\boldsymbol{q},\boldsymbol{k},\boldsymbol{p}\right)\equiv\int\!d^{2}\boldsymbol{r}\,e^{-i\left(\boldsymbol{q}\cdot\boldsymbol{r}\right)}\boldsymbol{r}^{i}\boldsymbol{r}^{j}\mathcal{H}\left(\boldsymbol{r}\right)\frac{e^{i\left(\boldsymbol{k}\cdot\boldsymbol{r}\right)}\left(e^{i\left(\boldsymbol{p}\cdot\boldsymbol{r}\right)}-1\right)}{\left(\boldsymbol{p}\cdot\boldsymbol{r}\right)\left(\boldsymbol{k}\cdot\boldsymbol{r}\right)},\label{eq:Ints2}
\end{equation}
so that the 1-gluon amplitude given in Eq. \eqref{eq:1Gfin} can be written as 
\begin{align}
\mathcal{A}_{1} & =\left(2\pi\right)\delta\left(p_{1}^{+}+p_{2}^{+}-p_{0}^{+}\right)\!\int\!d^{2}\boldsymbol{b}\,e^{-i\left(\boldsymbol{k}\cdot\boldsymbol{b}\right)}\label{eq:1Gfincomp}\\
 & \times\left[\bar{z}\,\mathcal{I}_{\mathcal{H}}^{i}\left(\boldsymbol{q},\bar{z}\boldsymbol{k}\right)\left(\partial^{i}U_{\boldsymbol{b}}^{R_{1}}\right)T^{R_{0}}U_{\boldsymbol{b}}^{R_{2}}-z\,\mathcal{I}_{\mathcal{H}}^{i}\left(\boldsymbol{q},-z\boldsymbol{k}\right)U_{\boldsymbol{b}}^{R_{1}}T^{R_{0}}\left(\partial^{i}U_{\boldsymbol{b}}^{R_{2}}\right)\right],\nonumber 
\end{align}
and the 2-gluon amplitude given in Eq. \eqref{eq:2Gfin} can be written as 
\begin{align}
\mathcal{A}_{2} & =\left(2\pi\right)\delta\left(p_{1}^{+}+p_{2}^{+}-p_{0}^{+}\right)\!\int\!\frac{d^{2}\boldsymbol{k}_{1}}{\left(2\pi\right)^{2}}\frac{d^{2}\boldsymbol{k}_{2}}{\left(2\pi\right)^{2}}\left(2\pi\right)^{2}\delta^{2}\left(\boldsymbol{k}_{1}+\boldsymbol{k}_{2}-\boldsymbol{k}\right)\nonumber \\
 & \times\int\!d^{2}\boldsymbol{b}_{1}d^{2}\boldsymbol{b}_{2}\,e^{-i\left(\boldsymbol{k}_{1}\cdot\boldsymbol{b}_{1}\right)-i\left(\boldsymbol{k}_{2}\cdot\boldsymbol{b}_{2}\right)}\left(\partial^{i}U_{\boldsymbol{b}_{1}}^{R_{1}}\right)T^{R_{0}}\left(\partial^{j}U_{\boldsymbol{b}_{2}}^{R_{2}}\right)\label{eq:2Gfincomp}\\
 & \times\left[z\mathcal{J}_{\mathcal{H}}^{ij}\left(\boldsymbol{q},-z\boldsymbol{k},\boldsymbol{k}_{1}\right)+\bar{z}\mathcal{J}_{\mathcal{H}}^{ij}\left(\boldsymbol{q},\bar{z}\boldsymbol{k},-\boldsymbol{k}_{2}\right)\right].\nonumber 
\end{align}

\section{Inclusive cross sections\label{sec:InclusiveXS}}

It is easy to obtain inclusive cross sections from our amplitudes given in Eq.~(\ref{eq:1Gfin})
and in Eq..~(\ref{eq:2Gfin}). In order to account for the possible use of
our results in the hybrid factorization ansatz, we average over
incoming projectile color states, with an averaging factor $C_{0}=N_{c}$
for a quark, $C_{0}=N_{c}^{2}-1$ for a gluon, and $C_{0}=1$ for
a photon. We also use the rapidities $y_1$ and $y_2$ of the outgoing particles. We can distinguish four contributions, depending on the number
of gluons in the TMD half-operator in the amplitude and in the complex
conjugate amplitude.

\subsection{TMD cross sections\label{subsec:TMDXS}}

\begin{figure}[h]
\begin{centering}
\includegraphics[width=0.9\textwidth]{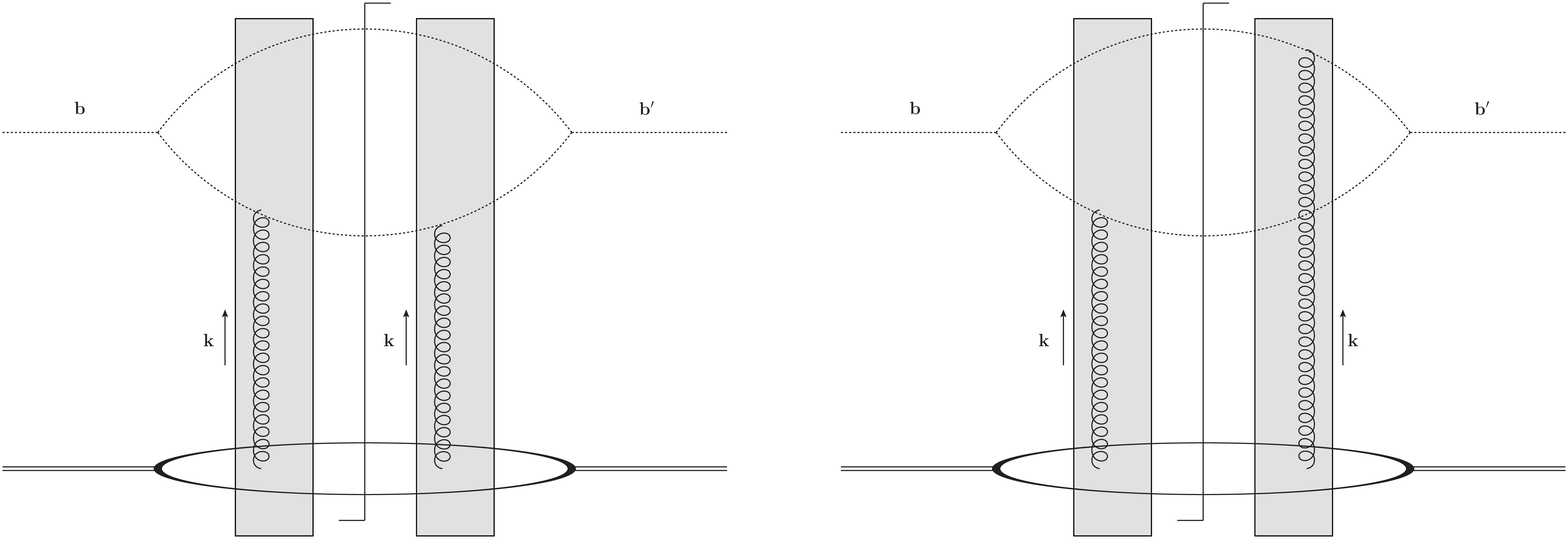}\vspace{0.5cm}
\includegraphics[width=0.9\textwidth]{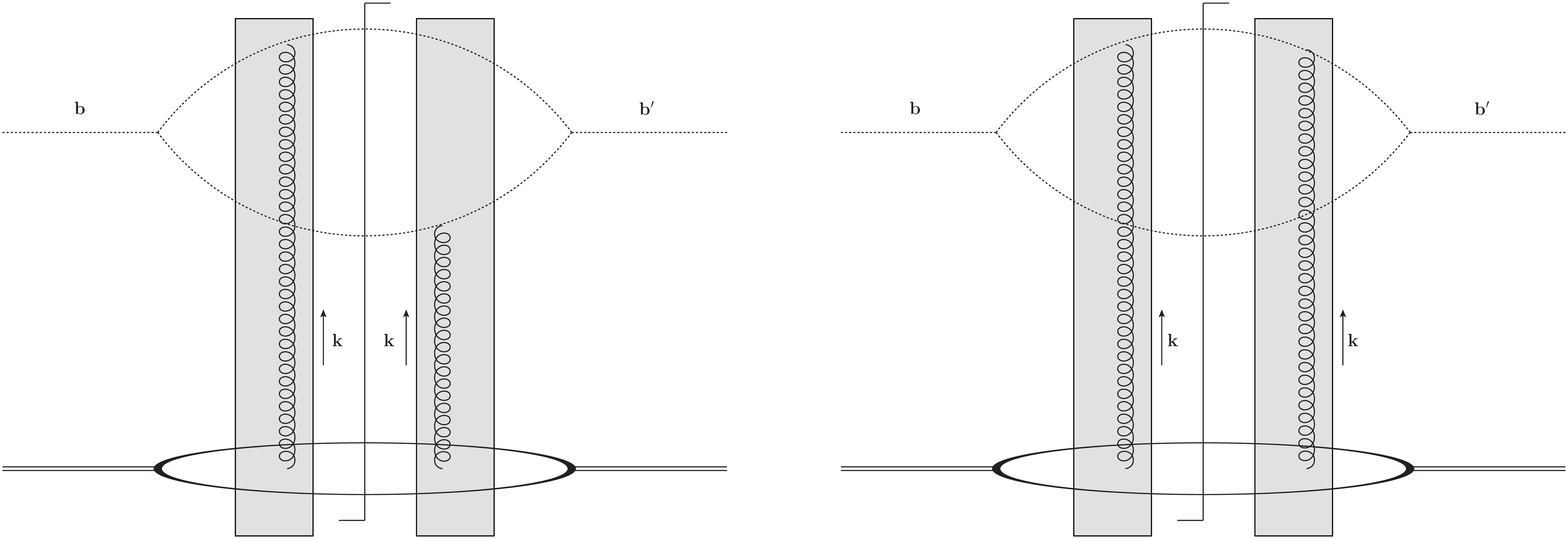}
\par\end{centering}
\caption{2-body contributions to the inclusive cross section. The gray blobs
represent interactions with low $k^{+}$ gluons via Wilson lines,
and the gluon line is isolated from the gauge link contributions by
differenciation of a Wilson line.\label{fig:sigma11}}
\end{figure}

\begin{figure}[h]
\centering{}\includegraphics[width=0.9\textwidth]{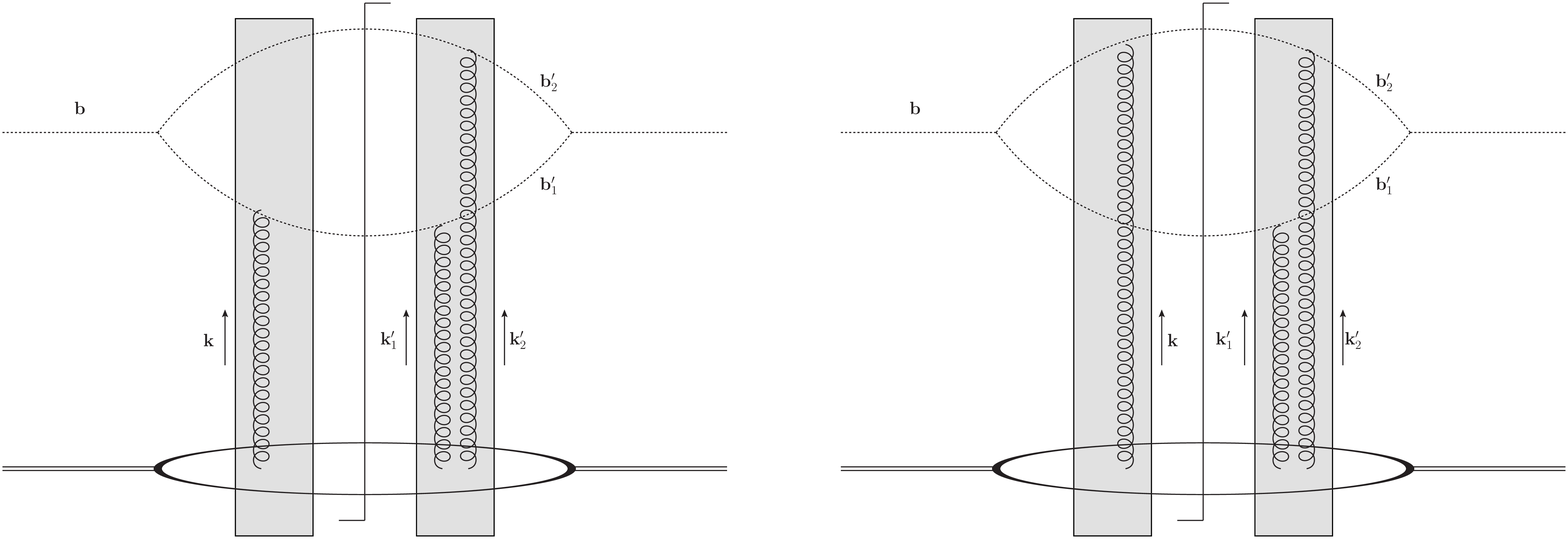}\caption{3-body contributions with 2 gluons in the complex conjugate amplitude\label{fig:sigma12}}
\end{figure}

The 2-body contribution, given by the four diagrams in Figure~\ref{fig:sigma11},
reads:
\begin{align}
 & \frac{d\sigma_{11}}{dy_1dy_2d^{2}\boldsymbol{q}d^{2}\boldsymbol{k}}=\frac{\delta\left(p_1^+ +p_2^+ -p_0^+\right)}{8\left(2\pi\right)C_{0}p_{0}^{+}}\int\!\frac{d^{2}\boldsymbol{b}^{\prime}}{\left(2\pi\right)^{2}}\frac{d^{2}\boldsymbol{b}}{\left(2\pi\right)^{2}}e^{i\boldsymbol{k}\cdot\left(\boldsymbol{b}^{\prime}-\boldsymbol{b}\right)}\label{eq:Incl11}\\
 & \times\left[\bar{z}^{2}\,\mathcal{I}_{\mathcal{H}}^{i}\left(\boldsymbol{q},\bar{z}\boldsymbol{k}\right)\mathcal{I}_{\mathcal{H}}^{j\ast}\left(\boldsymbol{q},\bar{z}\boldsymbol{k}\right)\frac{\left\langle P\left|\,\mathrm{Tr}\!\left[\left(\partial^{i}U_{\boldsymbol{b}}^{R_{1}}\right)T^{R_{0}}U_{\boldsymbol{b}}^{R_{2}}U_{\boldsymbol{b}^{\prime}}^{R_{2}\dagger}T^{R_{0}\dagger}\left(\partial^{j}U_{\boldsymbol{b}^{\prime}}^{R_{1}\dagger}\right)\right]\right|P\right\rangle }{\left\langle P|P\right\rangle }\right.\nonumber \\
 & -z\bar{z}\,\mathcal{I}_{\mathcal{H}}^{i}\left(\boldsymbol{q},\bar{z}\boldsymbol{k}\right)\mathcal{I}_{\mathcal{H}}^{j\ast}\left(\boldsymbol{q},-z\boldsymbol{k}\right)\frac{\left\langle P\left|\,\mathrm{Tr}\!\left[\left(\partial^{i}U_{\boldsymbol{b}}^{R_{1}}\right)T^{R_{0}}U_{\boldsymbol{b}}^{R_{2}}\left(\partial^{j}U_{\boldsymbol{b}^{\prime}}^{R_{2}\dagger}\right)T^{R_{0}\dagger}U_{\boldsymbol{b}^{\prime}}^{R_{1}\dagger}\right]\right|P\right\rangle }{\left\langle P|P\right\rangle }\nonumber \\
 & -z\bar{z}\,\mathcal{I}_{\mathcal{H}}^{i}\left(\boldsymbol{q},-z\boldsymbol{k}\right)\mathcal{I}_{\mathcal{H}}^{j\ast}\left(\boldsymbol{q},\bar{z}\boldsymbol{k}\right)\frac{\left\langle P\left|\,\mathrm{Tr}\!\left[U_{\boldsymbol{b}}^{R_{1}}T^{R_{0}}\left(\partial^{i}U_{\boldsymbol{b}}^{R_{2}}\right)U_{\boldsymbol{b}^{\prime}}^{R_{2}\dagger}T^{R_{0}\dagger}\left(\partial^{j}U_{\boldsymbol{b}^{\prime}}^{R_{1}\dagger}\right)\right]\right|P\right\rangle }{\left\langle P|P\right\rangle }\nonumber \\
 & \left.+z^{2}\,\mathcal{I}_{\mathcal{H}}^{i}\left(\boldsymbol{q},-z\boldsymbol{k}\right)\mathcal{I}_{\mathcal{H}}^{j\ast}\left(\boldsymbol{q},-z\boldsymbol{k}\right)\frac{\left\langle P\left|\,\mathrm{Tr}\!\left[U_{\boldsymbol{b}}^{R_{1}}T^{R_{0}}\left(\partial^{i}U_{\boldsymbol{b}}^{R_{2}}\right)\left(\partial^{j}U_{\boldsymbol{b}^{\prime}}^{R_{2}\dagger}\right)T^{R_{0}\dagger}U_{\boldsymbol{b}^{\prime}}^{R_{1}\dagger}\right]\right|P\right\rangle }{\left\langle P|P\right\rangle }\right],\nonumber 
\end{align}
the 3-body contributions are given by the diagrams with one gluon in
the amplitude and two gluons in the complex conjugate amplitude, as
in Figure~\ref{fig:sigma12}, which add up to:
\begin{align}
\frac{d\sigma_{12}}{dy_1dy_2d^{2}\boldsymbol{q}d^{2}\boldsymbol{k}} & =\frac{\delta\left(p_1^++p_2^+-p_0^+\right)}{8\left(2\pi\right)C_{0}p_{0}^{+}}\int\!\frac{d^{2}\boldsymbol{k}_{1}^{\prime}}{\left(2\pi\right)^{2}}\frac{d^{2}\boldsymbol{k}_{2}^{\prime}}{\left(2\pi\right)^{2}}\left(2\pi\right)^{2}\delta^{2}\left(\boldsymbol{k}_{1}^{\prime}+\boldsymbol{k}_{2}^{\prime}-\boldsymbol{k}\right)\nonumber \\
 & \times\int\!\frac{d^{2}\boldsymbol{b}}{\left(2\pi\right)^{2}}\frac{d^{2}\boldsymbol{b}_{1}^{\prime}d^{2}\boldsymbol{b}_{2}^{\prime}}{\left(2\pi\right)^{2}}e^{-i\left(\boldsymbol{k}\cdot\boldsymbol{b}\right)+i\left(\boldsymbol{k}_{1}^{\prime}\cdot\boldsymbol{b}_{1}^{\prime}\right)+i\left(\boldsymbol{k}_{2}^{\prime}\cdot\boldsymbol{b}_{2}^{\prime}\right)}\nonumber \\
 & \times\left[z\mathcal{J}_{\mathcal{H}}^{kl\ast}\left(\boldsymbol{q},-z\boldsymbol{k},\boldsymbol{k}_{1}^{\prime}\right)+\bar{z}\mathcal{J}_{\mathcal{H}}^{kl\ast}\left(\boldsymbol{q},\bar{z}\boldsymbol{k},-\boldsymbol{k}_{2}^{\prime}\right)\right]\label{eq:Incl12}\\
 & \times\left[\bar{z}\,\mathcal{I}_{\mathcal{H}}^{i}\left(\boldsymbol{q},\bar{z}\boldsymbol{k}\right)\frac{\left\langle P\left|\,\mathrm{Tr}\!\left[\left(\partial^{i}U_{\boldsymbol{b}}^{R_{1}}\right)T^{R_{0}}U_{\boldsymbol{b}}^{R_{2}}\left(\partial^{l}U_{\boldsymbol{b}_{2}^{\prime}}^{R_{2}\dagger}\right)T^{R_{0}\dagger}\left(\partial^{k}U_{\boldsymbol{b}_{1}^{\prime}}^{R_{1}\dagger}\right)\right]\right|P\right\rangle }{\left\langle P|P\right\rangle }\right.\nonumber \\
 & \left.-z\,\mathcal{I}_{\mathcal{H}}^{i}\left(\boldsymbol{q},-z\boldsymbol{k}\right)\frac{\left\langle P\left|\,\mathrm{Tr}\!\left[U_{\boldsymbol{b}}^{R_{1}}T^{R_{0}}\left(\partial^{i}U_{\boldsymbol{b}}^{R_{2}}\right)\left(\partial^{l}U_{\boldsymbol{b}_{2}^{\prime}}^{R_{2}\dagger}\right)T^{R_{0}\dagger}\left(\partial^{k}U_{\boldsymbol{b}_{1}^{\prime}}^{R_{1}\dagger}\right)\right]\right|P\right\rangle }{\left\langle P|P\right\rangle }\right],\nonumber 
\end{align}

\begin{figure}[h]
\centering{}\includegraphics[width=0.9\textwidth]{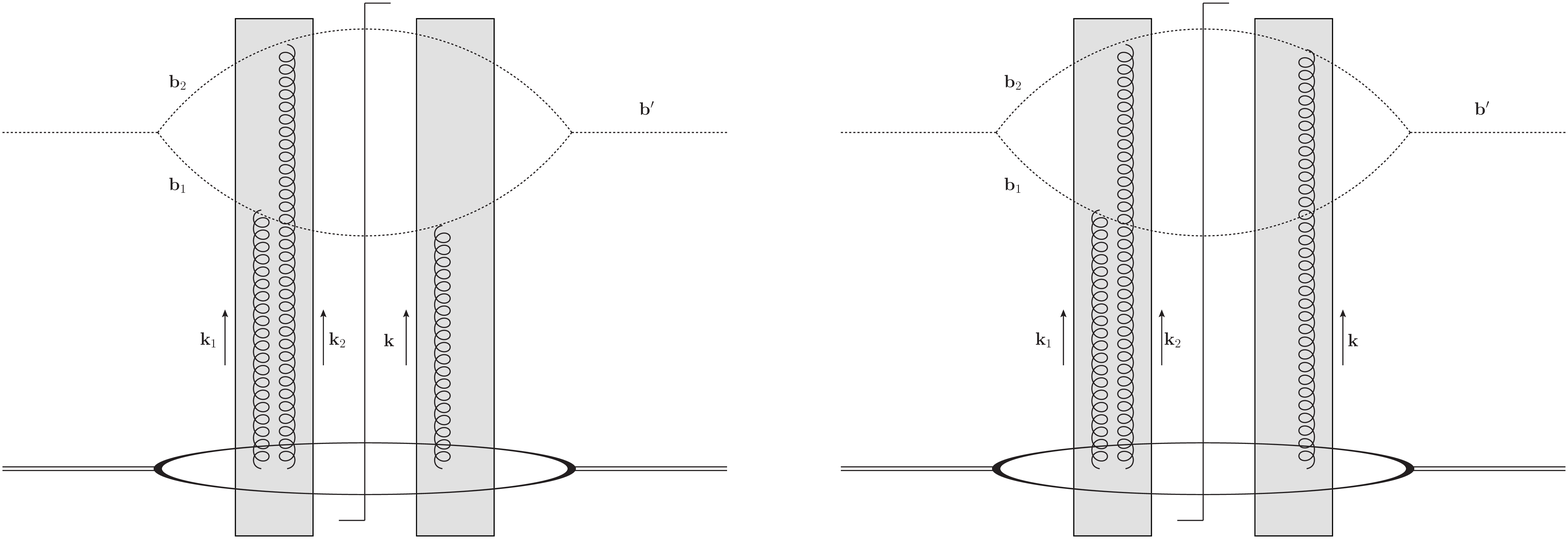}\caption{3-body contributions with 2 gluons in the amplitude\label{fig:sigma21}}
\end{figure}

and by those with two gluons in the amplitude and one in the complex conjugate
amplitude as in Figure~\ref{fig:sigma21}, which yield:
\begin{align}
\frac{d\sigma_{21}}{dy_1dy_2d^{2}\boldsymbol{q}d^{2}\boldsymbol{k}} & =\frac{\delta\left(p_1^++p_2^+-p_0^+\right)}{8\left(2\pi\right)C_{0}p_{0}^{+}}\int\!\frac{d^{2}\boldsymbol{k}_{1}}{\left(2\pi\right)^{2}}\frac{d^{2}\boldsymbol{k}_{2}}{\left(2\pi\right)^{2}}\left(2\pi\right)^{2}\delta^{2}\left(\boldsymbol{k}_{1}+\boldsymbol{k}_{2}-\boldsymbol{k}\right)\nonumber \\
 & \times\int\!\frac{d^{2}\boldsymbol{b}^{\prime}}{\left(2\pi\right)^{2}}\frac{d^{2}\boldsymbol{b}_{1}d^{2}\boldsymbol{b}_{2}}{\left(2\pi\right)^{2}}e^{i\left(\boldsymbol{k}\cdot\boldsymbol{b}^{\prime}\right)-i\left(\boldsymbol{k}_{1}\cdot\boldsymbol{b}_{1}\right)-i\left(\boldsymbol{k}_{2}\cdot\boldsymbol{b}_{2}\right)}\label{eq:Incl21}\\
 & \times\left[z\mathcal{J}_{\mathcal{H}}^{ij}\left(\boldsymbol{q},-z\boldsymbol{k},\boldsymbol{k}_{1}\right)+\bar{z}\mathcal{J}_{\mathcal{H}}^{ij}\left(\boldsymbol{q},\bar{z}\boldsymbol{k},-\boldsymbol{k}_{2}\right)\right]\nonumber \\
 & \times\left[\bar{z}\,\mathcal{I}_{\mathcal{H}}^{k\ast}\left(\boldsymbol{q},\bar{z}\boldsymbol{k}\right)\frac{\left\langle P\left|\,\mathrm{Tr}\!\left[\left(\partial^{i}U_{\boldsymbol{b}_{1}}^{R_{1}}\right)T^{R_{0}}\left(\partial^{j}U_{\boldsymbol{b}_{2}}^{R_{2}}\right)U_{\boldsymbol{b}^{\prime}}^{R_{2}\dagger}T^{R_{0}\dagger}\left(\partial^{k}U_{\boldsymbol{b}^{\prime}}^{R_{1}\dagger}\right)\right]\right|P\right\rangle }{\left\langle P|P\right\rangle }\right.\nonumber \\
 & \left.-z\,\mathcal{I}_{\mathcal{H}}^{k\ast}\left(\boldsymbol{q},-z\boldsymbol{k}\right)\frac{\left\langle P\left|\,\mathrm{Tr}\!\left[\left(\partial^{i}U_{\boldsymbol{b}_{1}}^{R_{1}}\right)T^{R_{0}}\left(\partial^{j}U_{\boldsymbol{b}_{2}}^{R_{2}}\right)\left(\partial^{k}U_{\boldsymbol{b}^{\prime}}^{R_{2}\dagger}\right)T^{R_{0}\dagger}U_{\boldsymbol{b}^{\prime}}^{R_{1}\dagger}\right]\right|P\right\rangle }{\left\langle P|P\right\rangle }\right].\nonumber 
\end{align}

\begin{figure}[h]
\begin{centering}
\includegraphics[width=0.5\textwidth]{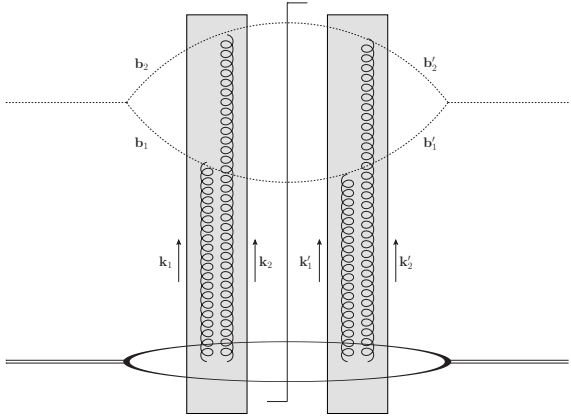}\caption{4-body contribution\label{fig:sigma22}}
\par\end{centering}
\end{figure}

Finally the 4-body contribution from the diagram in Figure~\ref{fig:sigma22},
reads
\begin{align}
 & \frac{d\sigma_{22}}{dy_1dy_2d^{2}\boldsymbol{q}d^{2}\boldsymbol{k}}=\frac{\delta\left(p_1^++p_2^+-p_0^+\right)}{8\left(2\pi\right)C_{0}p_{0}^{+}}\int\!\frac{d^{2}\boldsymbol{k}_{1}}{\left(2\pi\right)^{2}}\frac{d^{2}\boldsymbol{k}_{2}}{\left(2\pi\right)^{2}}\left(2\pi\right)^{2}\delta^{2}\left(\boldsymbol{k}_{1}+\boldsymbol{k}_{2}-\boldsymbol{k}\right)\nonumber \\
 & \times\int\!\frac{d^{2}\boldsymbol{k}_{1}^{\prime}}{\left(2\pi\right)^{2}}\frac{d^{2}\boldsymbol{k}_{2}^{\prime}}{\left(2\pi\right)^{2}}\left(2\pi\right)^{2}\delta^{2}\left(\boldsymbol{k}_{1}^{\prime}+\boldsymbol{k}_{2}^{\prime}-\boldsymbol{k}\right)\label{eq:Incl22}\\
 & \times\int\!\frac{d^{2}\boldsymbol{b}_{1}d^{2}\boldsymbol{b}_{2}}{\left(2\pi\right)^{2}}\frac{d^{2}\boldsymbol{b}_{1}^{\prime}d^{2}\boldsymbol{b}_{2}^{\prime}}{\left(2\pi\right)^{2}}e^{i\left(\boldsymbol{k}_{1}^{\prime}\cdot\boldsymbol{b}_{1}^{\prime}\right)+i\left(\boldsymbol{k}_{2}^{\prime}\cdot\boldsymbol{b}_{2}^{\prime}\right)-i\left(\boldsymbol{k}_{1}\cdot\boldsymbol{b}_{1}\right)-i\left(\boldsymbol{k}_{2}\cdot\boldsymbol{b}_{2}\right)}\nonumber \\
 & \times\left[z\mathcal{J}_{\mathcal{H}}^{ij}\left(\boldsymbol{q},-z\boldsymbol{k},\boldsymbol{k}_{1}\right)+\bar{z}\mathcal{J}_{\mathcal{H}}^{ij}\left(\boldsymbol{q},\bar{z}\boldsymbol{k},-\boldsymbol{k}_{2}\right)\right]\left[z\mathcal{J}_{\mathcal{H}}^{kl\ast}\left(\boldsymbol{q},-z\boldsymbol{k},\boldsymbol{k}_{1}^{\prime}\right)+\bar{z}\mathcal{J}_{\mathcal{H}}^{kl\ast}\left(\boldsymbol{q},\bar{z}\boldsymbol{k},-\boldsymbol{k}_{2}^{\prime}\right)\right]\nonumber \\
 & \times\frac{\left\langle P\left|\,\mathrm{Tr}\!\left[\left(\partial^{i}U_{\boldsymbol{b}_{1}}^{R_{1}}\right)T^{R_{0}}\left(\partial^{j}U_{\boldsymbol{b}_{2}}^{R_{2}}\right)\left(\partial^{l}U_{\boldsymbol{b}_{2}^{\prime}}^{R_{2}\dagger}\right)T^{R_{0}\dagger}\left(\partial^{k}U_{\boldsymbol{b}_{1}^{\prime}}^{R_{1}\dagger}\right)\right]\right|P\right\rangle }{\left\langle P|P\right\rangle }.\nonumber 
\end{align}
The inclusive (or incoherent) diffractive case is very similar to
the fully inclusive case. The difference lies in the TMD operators.
While the fully inclusive cross section involves $\langle P|\mathrm{tr}(\mathcal{O}_{\boldsymbol{x}}\mathcal{O}_{\boldsymbol{y}}^{\dagger})|P\rangle,$
the inclusive diffractive cross section involves $\langle P|\mathrm{tr}(\mathcal{O}_{\boldsymbol{x}}^{\left(1\right)}\mathcal{O}_{\boldsymbol{y}}^{\left(1\right)\dagger})|P\rangle,$
where $\mathcal{O}_{\boldsymbol{x}}^{\left(1\right)}$ and $\mathcal{O}_{\boldsymbol{y}}^{\left(1\right)\dagger}$
are the color singlet projections of the operators. In the CGC and
dipole descriptions of low $x$ physics, this matrix element is often
described as the $b$-dependent dipole scattering amplitude $\mathcal{N}(\boldsymbol{b},\boldsymbol{r})$. It is important to
note that the $\boldsymbol{b}$ variable which appears in these matrix
elements is the Fourier conjugate to the partonic transverse momentum
in a TMD. As a result, it should not be interpreted as the physical
impact parameter, which is the Fourier conjugate to the transverse
momentum imbalance in incoming and outgoing target states in a GTMD
or in a GPD. Instead, the $\boldsymbol{b}$ variable that appear in inclusive
diffraction is actually the transverse coordinate variable involved
in the Collins-Soper equation. This remark does not invalidate the
description of the b-dependent dipole scattering amplitude,
but it is important to keep in mind the nature of $\boldsymbol{b}$
when interpreting this quantity for inclusive observables.

\subsection{Cross sections with a PDF\label{subsec:PDFXS}}

A parton distribution function (PDF) is the integral of a TMD w.r.t.
its partonic transverse momenta. In order to obtain a cross section with a
PDF instead of a TMD, one should consider an inclusive process where
momentum $\boldsymbol{k}$ is not measured, and perform an expansion Eqs.~(\ref{eq:Incl11}, \ref{eq:Incl12},
\ref{eq:Incl21}, \ref{eq:Incl22}) in twists, by taking the leading
power in the hard scale. This hard scale $Q$ can be given by a virtuality or 
an invariant mass and thus the expansion is process-dependent. 
However, we can easily see how the leading kinematic twist part of
the leading genuine twist cross section, Eq.~(\ref{eq:Incl11}), can be
rewritten with a PDF: one considers $\left|\boldsymbol{k}\right|\ll Q$
in the hard factors $\mathcal{I}_{\mathcal{H}}^{i}$ then integrates
the cross section w.r.t. $\left|\boldsymbol{k}\right|$. For example
for a photon-induced process the leading twist cross section becomes
\begin{align}
\int\!d^{2}\boldsymbol{k}\left(\frac{d\sigma_{11}}{dy_1dy_2d^{2}\boldsymbol{q}d^{2}\boldsymbol{k}}\right)_{LT} & =\frac{\alpha_{s}\delta\left(p_1^++p_2^+-p_0^+\right)}{4C_{0}p_{0}^{+}}\mathcal{I}_{\mathcal{H}}^{i}\left(\boldsymbol{q},\boldsymbol{0}\right)\mathcal{I}_{\mathcal{H}}^{j\ast}\left(\boldsymbol{q},\boldsymbol{0}\right)\label{eq:PDFxs}\\
 & \times\int\!db^{+}\left.\frac{\left\langle P\left|\,\mathrm{Tr}\!\left[F^{i-}\left(b\right)F^{j-}\left(0\right)\right]\right|P\right\rangle }{2P^{-}\left(2\pi\right)^{2}}\right|_{b^{-}=0,\boldsymbol{b}=\boldsymbol{0}},\nonumber 
\end{align}
where one can easily identify a PDF in the $x\sim0$ approximation in
the second line.

\section{Exclusive cross sections\label{sec:ExclusiveXS}}

\subsection{GTMD cross sections\label{subsec:GTMDXS}}

For exclusive cross sections, only the 2-body amplitudes contribute since the target matrix elements in this case is a color singlet operator. The off-diagonal matrix elements of the 2-body operators can easily
be identified as GTMDs. Here, we denote the momentum imbalance as $\Delta_\perp$ rather than $k_\perp$ to match more standard notations for the GTMD. The exclusive cross section reads
\begin{align}
 & \frac{d\sigma_{excl}}{dy_1dy_2d^{2}\boldsymbol{q}d^{2}\boldsymbol{\Delta}}=\frac{\delta\left(p_1^++p_2^+-p_0^+\right)}{8\left(2\pi\right)C_{0}p_{0}^{+}}\int\frac{d^{2}\boldsymbol{k}_{1}}{\left(2\pi\right)^{2}}\frac{d^{2}\boldsymbol{k}_{2}}{\left(2\pi\right)^{2}}\left(2\pi\right)^{2}\delta^{2}\left(\boldsymbol{k}_{1}+\boldsymbol{k}_{2}-\boldsymbol{\Delta}\right)\nonumber \\
 & \times\int\frac{d^{2}\boldsymbol{k}_{1}^{\prime}}{\left(2\pi\right)^{2}}\frac{d^{2}\boldsymbol{k}_{2}^{\prime}}{\left(2\pi\right)^{2}}\left(2\pi\right)^{2}\delta^{2}\left(\boldsymbol{k}_{1}^{\prime}+\boldsymbol{k}_{2}^{\prime}-\boldsymbol{\Delta}\right)\nonumber \\
 & \times\int\frac{d^{2}\boldsymbol{b}_{1}d^{2}\boldsymbol{b}_{2}}{\left(2\pi\right)^{2}}\frac{d^{2}\boldsymbol{b}_{1}^{\prime}d^{2}\boldsymbol{b}_{2}^{\prime}}{\left(2\pi\right)^{2}}e^{i\left(\boldsymbol{k}_{1}^{\prime}\cdot\boldsymbol{b}_{1}^{\prime}\right)+i\left(\boldsymbol{k}_{2}^{\prime}\cdot\boldsymbol{b}_{2}^{\prime}\right)-i\left(\boldsymbol{k}_{1}\cdot\boldsymbol{b}_{1}\right)-i\left(\boldsymbol{k}_{2}\cdot\boldsymbol{b}_{2}\right)}\label{eq:Excl}\\
 & \times\left[z\mathcal{J}_{\mathcal{H}}^{ij}\left(\boldsymbol{q},-z\boldsymbol{\Delta},\boldsymbol{k}_{1}\right)+\bar{z}\mathcal{J}_{\mathcal{H}}^{ij}\left(\boldsymbol{q},\bar{z}\boldsymbol{\Delta},-\boldsymbol{k}_{2}\right)\right]\left[z\mathcal{J}_{\mathcal{H}}^{kl\ast}\left(\boldsymbol{q},-z\boldsymbol{\Delta},\boldsymbol{k}_{1}^{\prime}\right)+\bar{z}\mathcal{J}_{\mathcal{H}}^{kl\ast}\left(\boldsymbol{q},\bar{z}\boldsymbol{\Delta},-\boldsymbol{k}_{2}^{\prime}\right)\right]\nonumber \\
 & \times\mathrm{tr}_{c}\frac{\Bigl\langle P-\boldsymbol{\Delta}\Bigl|\left[\left(\partial^{i}U_{\boldsymbol{b}_{1}}^{R_{1}}\right)T^{R_{0}}\left(\partial^{j}U_{\boldsymbol{b}_{2}}^{R_{2}}\right)\right]^{\left(1\right)}\Bigr|P\Bigr\rangle}{\left\langle P|P\right\rangle }\frac{\Bigl\langle P\Bigl|\left[\left(\partial^{l}U_{\boldsymbol{b}_{2}^{\prime}}^{R_{2}\dagger}\right)T^{R_{0}\dagger}\left(\partial^{k}U_{\boldsymbol{b}_{1}^{\prime}}^{R_{1}\dagger}\right)\right]^{\left(1\right)}\Bigr|P-\boldsymbol{\Delta}\Bigr\rangle}{\left\langle P|P\right\rangle },\nonumber 
\end{align}
where $\mathrm{tr}_{c}$ is the trace over all remaining open color
indices in the product of distributions. For example in a $g\rightarrow q\bar{q}$
cross section the last line in Eq.~(\ref{eq:Excl}) would read
\begin{equation}
\delta^{ab}\frac{\left\langle P-\boldsymbol{\Delta}\left|\frac{1}{2}\mathrm{Tr}\left[\left(\partial^{i}U_{\boldsymbol{b}_{1}}\right)T^{a}\left(\partial^{j}U_{\boldsymbol{b}_{2}}^{\dagger}\right)\right]\right|P\right\rangle }{\left\langle P|P\right\rangle }\frac{\left\langle P\left|\frac{1}{2}\mathrm{Tr}\left[\left(\partial^{l}U_{\boldsymbol{b}_{2}^{\prime}}\right)T^{b\dagger}\left(\partial^{k}U_{\boldsymbol{b}_{1}^{\prime}}^{\dagger}\right)\right]\right|P-\boldsymbol{\Delta}\right\rangle }{\left\langle P|P\right\rangle }.\label{eq:gqq}
\end{equation}
The non-perturbative matrix elements involved in Eq.~(\ref{eq:Excl})
are GTMDs. We would like to emphasize the fact that Eq.~(\ref{eq:Excl})
is exact. Here, it shows a perfect match between exclusive low $x$
cross sections and twist-resummed GTMD cross sections in the small
$x$ limit.

\subsection{Cross sections with a GPD\label{subsec:GPDXS}}

The GPD limit is obtained from a GTMD cross section the same way the
PDF limit is obtained from a TMD cross section, noting that a GPD
is the integral of a GTMD with respect to partonic transverse momenta. One
performs a twist expansion by taking $\boldsymbol{k}_{1,2}^{\left(\prime\right)}/Q\rightarrow0$\textbf{
}in the hard parts, then integrates over partonic transverse momenta.
For example for photon-induced processes at leading twist we get
\begin{align}
 & \left(\frac{d\sigma_{excl}^{GPD}}{dy_1dy_2d^{2}\boldsymbol{q}d^{2}\boldsymbol{\Delta}}\right)_{LT}=\frac{\alpha_{s}^{2}\delta\left(p_1^++p_2^+-p_0^+\right)}{8\left(2\pi\right)C_{0}p_{0}^{+}}\left[z\mathcal{J}_{\mathcal{H}}^{ij}\left(\boldsymbol{q},-z\boldsymbol{\Delta},\boldsymbol{0}\right)+\bar{z}\mathcal{J}_{\mathcal{H}}^{ij}\left(\boldsymbol{q},\bar{z}\boldsymbol{\Delta},\boldsymbol{0}\right)\right]\nonumber \\
 & \times\left[z\mathcal{J}_{\mathcal{H}}^{kl\ast}\left(\boldsymbol{q},-z\boldsymbol{\Delta},\boldsymbol{0}\right)+\bar{z}\mathcal{J}_{\mathcal{H}}^{kl\ast}\left(\boldsymbol{q},\bar{z}\boldsymbol{\Delta},\boldsymbol{0}\right)\right]\label{eq:GPDxs}\\
 & \times\int\!\frac{db^{+}}{2\pi P^{-}}\left.\left\langle P-\boldsymbol{\Delta}\left|\mathrm{Tr}\left[F^{i-}\left(b\right)\left[b^{+},0^{+}\right]_{\boldsymbol{0}}F^{j-}\left(0\right)\left[0^{+},b^{+}\right]_{\boldsymbol{0}}\right]\right|P\right\rangle \right|_{b^{-}=0,\boldsymbol{b}=\boldsymbol{0}}\nonumber \\
 & \times\int\!\frac{db^{\prime+}}{2\pi P^{-}}\left.\left\langle P\left|\mathrm{Tr}\left[F^{l-}\left(b^{\prime}\right)\left[b^{\prime+},0^{+}\right]_{\boldsymbol{0}}F^{k-}\left(0\right)\left[0^{+},b^{\prime+}\right]_{\boldsymbol{0}}\right]\right|P-\boldsymbol{\Delta}\right\rangle \right|_{b^{\prime-}=0,\boldsymbol{b}^{\prime}=\boldsymbol{0}}.\nonumber 
\end{align}
At leading twist the gauge links do not contribute, and one can easily
recognize leading twist GPDs in the last two lines.

\section{The BFKL limit as a kinematic limit\label{sec:BFKL}}

The BFKL limit is usually understood as a weak field limit $gF\sim0$,
known as the dilute limit. In a recent study~\citep{Altinoluk:2019fui},
it was shown how this limit could also be recovered by using
the Wandzura-Wilczek approximation in the CGC and identifying all
gluon distributions as the unintegrated PDF, which is justified at
large $\left|\boldsymbol{k}\right|$. In this section, we aim at describing
the BFKL limit as a kinematic limit rather than a weak field limit.
BFKL is valid when all transverse momenta are of the order of the
hard scale, and we are interested in studying  BFKL beyond the WW approximation,
so let us consider the limit of large
partonic transverse momenta. By Fourier conjugation, this limit leads
to the shrinking of transverse gauge links:
\begin{equation}
\left[x^{+},\pm\infty\right]_{\boldsymbol{b}_{i}}\left[\pm\infty,y^{+}\right]_{\boldsymbol{b}_{j}}\sim\left[x^{+},y^{+}\right]_{\boldsymbol{b}_{i}\sim\boldsymbol{b}_{j}\sim\boldsymbol{0}}.\label{eq:shrinking}
\end{equation}
This makes all gauge links unidimensional and in the same direction,
which means all 2-body distributions can be rewritten as
\begin{equation}
\int\!\frac{d^{2}\boldsymbol{k}}{\left(2\pi\right)^{2}}e^{-i\left(\boldsymbol{k}\cdot\boldsymbol{x}\right)}\int\!dx^{+}\left\langle P\left|F^{i-}\left(x\right)\left[x^{+},0^{+}\right]_{\boldsymbol{0}}F^{j-}\left(0\right)\left[0^{+},x^{+}\right]_{\boldsymbol{0}}\right|P\right\rangle \biggr|_{x^{-}=0}\label{eq:uPDF}
\end{equation}
since the modification of gauge links between $x$ and $0$ in the
transverse plane is free up to small corrections. This unique distribution
is the 2-body unintegrated PDF. In $A^{+}=0$ gauge, it can be rewritten
as
\begin{equation}
\int\!\frac{d^{2}\boldsymbol{k}}{\left(2\pi\right)^{2}}e^{-i\left(\boldsymbol{k}\cdot\boldsymbol{x}\right)}\frac{\boldsymbol{k}^{i}\boldsymbol{k}^{j}}{\boldsymbol{k}^{2}}\boldsymbol{k}^{2}\left\langle P\left|A^{-}\left(x\right)\left[x^{+},0^{+}\right]_{\boldsymbol{0}}A^{-}\left(0\right)\left[0^{+},x^{+}\right]_{\boldsymbol{0}}\right|P\right\rangle \biggr|_{x^{-}=0},\label{eq:uPDFLC}
\end{equation}
where one can explicitely identify the so-called \textit{nonsense polarization}  vector
in lightcone gauge $\frac{\boldsymbol{k}^{i}}{\left|\boldsymbol{k}\right|}$.\\
The importance of gauge links at small $k_\perp$ and the shrinking of all TMD distributions into the unique unintegrated PDF was observed and confirmed numerically in~\cite{Marquet:2016cgx, Marquet:2017xwy}. \\
Similarly to Eq.(~\ref{eq:uPDF}), all 3-body distributions become
\begin{align}
 & \int\frac{d^{2}\boldsymbol{k}_{1}}{\left(2\pi\right)^{2}}\frac{d^{2}\boldsymbol{k}_{2}}{\left(2\pi\right)^{2}}e^{-i\left(\boldsymbol{k}_{1}\cdot\boldsymbol{x}_{1}\right)-i\left(\boldsymbol{k}_{2}\cdot\boldsymbol{x}_{2}\right)}\int dx_{1}^{+}dx_{2}^{+}\label{eq:3BuPDF}\\
 & \times\left\langle P\left|F^{i-}\left(x_{1}\right)\left[x_{1}^{+},x_{2}^{+}\right]_{\boldsymbol{0}}g_{s}F^{j-}\left(x_{2}\right)\left[x_{2}^{+},0^{+}\right]_{\boldsymbol{0}}F^{k-}\left(0\right)\left[0^{+},x_{1}^{+}\right]_{\boldsymbol{0}}\right|P\right\rangle \biggr|_{x_{1,2}^{-}=0},\nonumber 
\end{align}
where it is important to keep $g_{s}$ in the operator rather than
the hard part. Indeed genuine twist corrections do not come with a
perturbative $g_{s}$ suppression: the $g_{s}$ factor is in the non-perturbative
matrix element, which means the 3-body contributions are of the same
order of perturbation theory as the 2-body contributions. In most of the studies which appear in the BFKL literature, the 3- and 4-body unintegrated PDFs are
not usually taken into account, with the ill-advised
assumption that their contributions are $\alpha_{s}$- suppressed.
Here, we observe that they are actually dropped as a Wandzura-Wilczek
approximation, which does not have a perturbative origin. Understanding
BFKL as a kinematic limit means that all genuine twist corrections
should be taken into account. For example for photon-induced processes
in the BFKL kinematic limit and in $A^+=0$ gauge, Eqs.~(\ref{eq:Incl11},
\ref{eq:Incl12}, \ref{eq:Incl21}, \ref{eq:Incl22}) become respectively
the 2-body contribution
\begin{align}
 & \frac{d\sigma_{11}}{dy_1dy_2d^{2}\boldsymbol{q}d^{2}\boldsymbol{k}}\sim\frac{\alpha_{s}\delta\left(p_1^++p_2^+-p_0^+\right)}{4\left(2\pi\right)C_{0}p_{0}^{+}}\frac{\boldsymbol{k}^{i}\boldsymbol{k}^{j}}{\boldsymbol{k}^{2}}\nonumber \\
 & \times\left[\bar{z}\,\mathcal{I}_{\mathcal{H}}^{i}\left(\boldsymbol{q},\bar{z}\boldsymbol{k}\right)+z\,\mathcal{I}_{\mathcal{H}}^{i}\left(\boldsymbol{q},-z\boldsymbol{k}\right)\right]\left[\bar{z}\,\mathcal{I}_{\mathcal{H}}^{j\ast}\left(\boldsymbol{q},\bar{z}\boldsymbol{k}\right)+z\,\mathcal{I}_{\mathcal{H}}^{j\ast}\left(\boldsymbol{q},-z\boldsymbol{k}\right)\right]\label{eq:BFKL11}\\
 & \times\frac{\boldsymbol{k}^{2}}{2P^{-}}\int\!\frac{d^{2}\boldsymbol{b}}{\left(2\pi\right)^{2}}e^{-i\left(\boldsymbol{k}\cdot\boldsymbol{b}\right)}\int\!\frac{db^{+}}{2\pi}\left\langle P\left|\mathrm{Tr}\left[A^{-}\left(b\right)\left[b^{+},0^{+}\right]_{\boldsymbol{0}}A^{-}\left(0\right)\left[0^{+},b^{+}\right]_{\boldsymbol{0}}\right]\right|P\right\rangle \biggr|_{b^{-}=0},\nonumber 
\end{align}
the 3-body contributions
\begin{align}
 & \frac{d\sigma_{12}}{dy_1dy_2d^{2}\boldsymbol{q}d^{2}\boldsymbol{k}}=\frac{\alpha_{s}\delta\left(p_1^++p_2^+-p_0^+\right)}{4C_{0}p_{0}^{+}}\int\!\frac{d^{2}\boldsymbol{k}_{1}^{\prime}}{\left(2\pi\right)^{2}}\frac{d^{2}\boldsymbol{k}_{2}^{\prime}}{\left(2\pi\right)^{2}}\left(2\pi\right)^{2}\delta^{2}\left(\boldsymbol{k}_{1}^{\prime}+\boldsymbol{k}_{2}^{\prime}-\boldsymbol{k}\right)\label{eq:BFKL12}\\
 & \times\left(\frac{\boldsymbol{k}^{i}\boldsymbol{k}_{1}^{\prime k}\boldsymbol{k}_{2}^{\prime l}}{\boldsymbol{k}^{2}}\right)\left[\bar{z}\,\mathcal{I}_{\mathcal{H}}^{i}\left(\boldsymbol{q},\bar{z}\boldsymbol{k}\right)+z\,\mathcal{I}_{\mathcal{H}}^{i}\left(\boldsymbol{q},-z\boldsymbol{k}\right)\right]\left[\bar{z}\mathcal{J}_{\mathcal{H}}^{kl\ast}\left(\boldsymbol{q},\bar{z}\boldsymbol{k},-\boldsymbol{k}_{2}^{\prime}\right)+z\mathcal{J}_{\mathcal{H}}^{kl\ast}\left(\boldsymbol{q},-z\boldsymbol{k},\boldsymbol{k}_{1}^{\prime}\right)\right]\nonumber \\
 & \times\frac{\boldsymbol{k}^{2}}{2P^{-}}\int\!\frac{d^{2}\boldsymbol{b}}{\left(2\pi\right)^{2}}\frac{d^{2}\boldsymbol{b}^{\prime}}{\left(2\pi\right)^{2}}e^{-i\left(\boldsymbol{k}\cdot\boldsymbol{b}\right)+i\left(\boldsymbol{k}_{2}^{\prime}\cdot\boldsymbol{b}^{\prime}\right)}\int\!db^{+}db^{\prime+}\nonumber \\
 & \times\left\langle P\left|\mathrm{Tr}\left[A^{-}\left(b\right)\left[b^{+},b^{\prime+}\right]_{\boldsymbol{0}}g_{s}A^{-}\left(b^{\prime}\right)\left[b^{\prime+},0^{+}\right]_{\boldsymbol{0}}A^{-}\left(0\right)\left[0^{+},b^{+}\right]_{\boldsymbol{0}}\right]\right|P\right\rangle \biggr|_{b^{\left(\prime\right)-}=0},\nonumber 
\end{align}
and 
\begin{align}
 & \frac{d\sigma_{21}}{dy_1dy_2d^{2}\boldsymbol{q}d^{2}\boldsymbol{k}}=\frac{\alpha_{s}\delta\left(p_1^++p_2^+-p_0^+\right)}{2C_{0}p_{0}^{+}}\int\!\frac{d^{2}\boldsymbol{k}_{1}}{\left(2\pi\right)^{2}}\frac{d^{2}\boldsymbol{k}_{2}}{\left(2\pi\right)^{2}}\left(2\pi\right)^{2}\delta^{2}\left(\boldsymbol{k}_{1}+\boldsymbol{k}_{2}-\boldsymbol{k}\right)\label{eq:BFKL21}\\
 & \times\left(\frac{\boldsymbol{k}_{1}^{i}\boldsymbol{k}_{2}^{j}\boldsymbol{k}^{k}}{\boldsymbol{k}^{2}}\right)\left[\bar{z}\mathcal{J}_{\mathcal{H}}^{ij}\left(\boldsymbol{q},\bar{z}\boldsymbol{k},-\boldsymbol{k}_{2}\right)+z\mathcal{J}_{\mathcal{H}}^{ij}\left(\boldsymbol{q},-z\boldsymbol{k},\boldsymbol{k}_{1}\right)\right]\left[\bar{z}\,\mathcal{I}_{\mathcal{H}}^{k\ast}\left(\boldsymbol{q},\bar{z}\boldsymbol{k}\right)+z\,\mathcal{I}_{\mathcal{H}}^{k\ast}\left(\boldsymbol{q},-z\boldsymbol{k}\right)\right]\nonumber \\
 & \times\frac{\boldsymbol{k}^{2}}{2P^{-}}\int\!\frac{d^{2}\boldsymbol{b}}{\left(2\pi\right)^{2}}\frac{d^{2}\boldsymbol{b}^{\prime}}{\left(2\pi\right)^{2}}e^{-i\left(\boldsymbol{k}_{1}\cdot\boldsymbol{b}\right)-i\left(\boldsymbol{k}_{2}\cdot\boldsymbol{b}^{\prime}\right)}\int\!db^{+}db^{\prime+}\nonumber \\
 & \times\left\langle P\left|\mathrm{Tr}\left[A^{-}\left(b\right)\left[b^{+},b^{\prime+}\right]_{\boldsymbol{0}}g_{s}A^{-}\left(b^{\prime}\right)\left[b^{\prime+},0^{+}\right]_{\boldsymbol{0}}A^{-}\left(0\right)\left[0^{+},b^{+}\right]_{\boldsymbol{0}}\right]\right|P\right\rangle \biggr|_{b^{\left(\prime\right)-}=0},\nonumber 
\end{align}
and finally the 4-body contribution 
\begin{align}
 & \frac{d\sigma_{22}}{dy_1dy_2d^{2}\boldsymbol{q}d^{2}\boldsymbol{k}}=\frac{\alpha_{s}\delta\left(p_1^++p_2^+-p_0^+\right)}{2C_{0}p_{0}^{+}}\int\!\frac{d^{2}\boldsymbol{k}_{1}}{\left(2\pi\right)^{2}}\frac{d^{2}\boldsymbol{k}_{2}}{\left(2\pi\right)^{2}}\left(2\pi\right)^{2}\delta\left(\boldsymbol{k}_{1}+\boldsymbol{k}_{2}-\boldsymbol{k}\right)\nonumber \\
 & \times\int\!\frac{d^{2}\boldsymbol{k}_{1}^{\prime}}{\left(2\pi\right)^{2}}\frac{d^{2}\boldsymbol{k}_{2}^{\prime}}{\left(2\pi\right)^{2}}\left(2\pi\right)^{2}\delta\left(\boldsymbol{k}_{1}^{\prime}+\boldsymbol{k}_{2}^{\prime}-\boldsymbol{k}\right)\label{eq:BFKL22}\\
 & \times\frac{\boldsymbol{k}_{1}^{i}\boldsymbol{k}_{2}^{j}}{\boldsymbol{k}^{2}}\left[z\mathcal{J}_{\mathcal{H}}^{ij}\left(\boldsymbol{q},-z\boldsymbol{k},\boldsymbol{k}_{1}\right)+\bar{z}\mathcal{J}_{\mathcal{H}}^{ij}\left(\boldsymbol{q},\bar{z}\boldsymbol{k},-\boldsymbol{k}_{2}\right)\right]\nonumber \\
 & \times\frac{\boldsymbol{k}_{1}^{\prime k}\boldsymbol{k}_{2}^{\prime l}}{\boldsymbol{k}^{2}}\left[z\mathcal{J}_{\mathcal{H}}^{kl\ast}\left(\boldsymbol{q},-z\boldsymbol{k},\boldsymbol{k}_{1}^{\prime}\right)+\bar{z}\mathcal{J}_{\mathcal{H}}^{kl\ast}\left(\boldsymbol{q},\bar{z}\boldsymbol{k},-\boldsymbol{k}_{2}^{\prime}\right)\right]\nonumber \\
 & \times\int\!\frac{d^{2}\boldsymbol{b}_{1}d^{2}\boldsymbol{b}_{2}}{\left(2\pi\right)^{2}}\frac{d^{2}\boldsymbol{b}^{\prime}}{\left(2\pi\right)^{2}}e^{i\left(\boldsymbol{k}_{2}^{\prime}\cdot\boldsymbol{b}^{\prime}\right)-i\left(\boldsymbol{k}_{1}\cdot\boldsymbol{b}_{1}\right)-i\left(\boldsymbol{k}_{2}\cdot\boldsymbol{b}_{2}\right)}\int\!db_{1}^{+}db_{2}^{+}db^{\prime+}\nonumber \\
 & \times\frac{\boldsymbol{k}^{4}}{2P^{-}}\Bigl\langle P\Bigl|\mathrm{Tr}\Bigl(A^{-}\left(b_{1}\right)\left[b_{1}^{+},b_{2}^{+}\right]_{\boldsymbol{0}}g_{s}A^{-}\left(b_{2}\right)\left[b_{2}^{+},b^{\prime+}\right]_{\boldsymbol{0}}\nonumber \\
 & \times g_{s}A^{-}\left(b^{\prime}\right)\left[b^{\prime+},0^{+}\right]_{\boldsymbol{0}}A^{-}\left(0\right)\left[0^{+},b_{1}^{+}\right]_{\boldsymbol{0}}\Bigr)\Bigr|P\Bigr\rangle\biggr|_{b_{1,2}^{-}=b^{\prime-}=0}.\nonumber 
\end{align}
We emphasize that all four contributions are of the same
order in perturbation theory, and neglecting the contributions from
genuine higher twist unintegrated PDFs is justified as a Wandzura-Wilczek
approximation rather than a perturbative suppression. The validity
of this approximation should be evaluated for each process.

\section{The origins of saturation\label{sec:Saturation}}

\begin{figure}[h]\begin{center}
\includegraphics[width=0.6\textwidth]{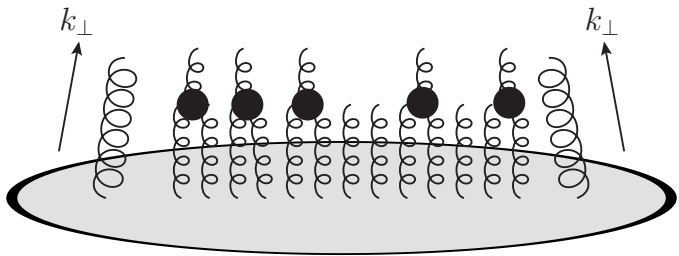} \caption{Saturation effects from the evolution: the non-linearities in the evolution equation account for recombination effects in the target\label{fig:evosat}}
\end{center}
\end{figure}

In our formulation of low $x$ physics, saturation can be understood
as three distinct effects. \\
First of all, a well known form of saturation is evolutional, and arises from the non-linearity of the B-JIMWLK hierarchy of evolution equations and its truncated and approximated daughter equations (see Figure~\ref{fig:evosat}). This
non-linearity is expected to slow down the growth in $s$ of low $x$
cross sections~\cite{Gribov:1984tu}, thus contributing to restoring the unitarity of the S matrix.  

\begin{figure}\begin{center}
\includegraphics[width=0.6\textwidth]{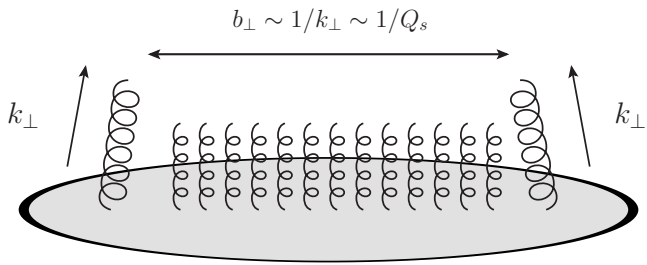} \caption{Kinematic saturation: the separation between the TMD gluons is filled by multiple soft scatterings. \label{fig:kinsat}}
\end{center}
\end{figure}

The other effects are due to multiple scattering via interactions
with slow gluons, but we distinguish two types of such effects.
We refer to the first type, described in Figure~\ref{fig:kinsat}, as the \textit{kinematic saturation}. It is linked to the gauge link structures of the gluon distributions. Indeed, the gauge links account for multiple scatterings from slow gluons, and the importance of such gauge links is that they can be used as a probe for multiple scattering effects. These effects have been investigated recently in~\cite{Marquet:2016cgx, Marquet:2017xwy}. They are expected to appear at small $\left|\boldsymbol{k}\right|$, since all TMD distributions reduce to the unintegrated PDF in the large $\left| \boldsymbol{k} \right|$ limit regardless of their gauge link structure, as discussed earlier. By studying the behavior of different distributions all along the
$\left|\boldsymbol{k}\right|$ range, it was shown that indeed distributions
with distinct gauge link structures have to be distinguished at low
$\left|\boldsymbol{k}\right|$ while at large $\left|\boldsymbol{k}\right|$
all distributions are the same. This kind of multiple scattering is thus due to the presence of a large transverse coordinate region, conjugate to $\left|\boldsymbol{k}\right|$, to fill with the soft gluons in that kinematic regime.  

\begin{figure}\begin{center}
\includegraphics[width=0.6\textwidth]{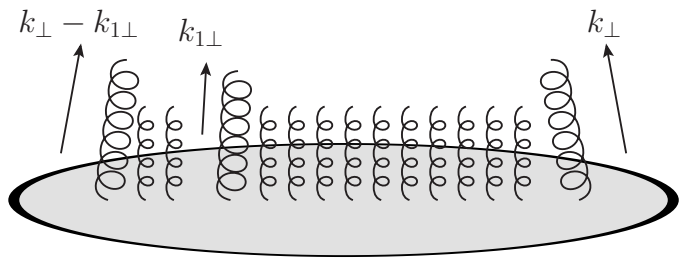} \caption{Genuine saturation: for dense targets where gluon occupancy is large, the probability to extract more gluons is enhanced, hence an expected enhancement of genuine twist corrections.\label{fig:gensat}}
\end{center}
\end{figure}

Finally the last type of saturation, described in Figure~\ref{fig:gensat}, to which we refer as \textit{genuine saturation}, is due to genuine twist corrections.
In addition to the gluons forming the gauge links, the extra gluons
from higher twist operators can contribute to multiple scattering
effects. Given that the genuine twist corrections in an operator are
obtained in physical gauges by the insertion of a gluon field along
with the coupling constant $g_{s}$ and the appropriate gauge links,
the genuine twist corrections are not perturbatively suppressed as
assumed implicitely in most studies involving unintegrated PDFs\footnote{Obviously this remark only concerns non-perturbative targets. BFKL
resummation is valid in full generality for soft gluon exchanges between
perturbative objects, for example Mueller-Navelet jets~\cite{Mueller:1986ey}.}: the $g_{s}$ factor is part of the non-perturbative matrix elements
and neglecting them is tantamount to using the Wandzura-Wilczek approximation.
This kind of saturation effects would appear even in the high $\left|\boldsymbol{k}\right|$
BFKL regime if one does not restrict oneself to this unquantified
approximation, whose validity should be tested in a process-dependent
way. In the CGC picture, the large gluon occupancy number in a dense target leads to the scaling $g_s F \sim 1$, which leads to an expected enhancement of genuine twist corrections. In that sense, genuine saturation can be understood as the invalidation of the Wandzura-Wilczek approximation.

\section{Discussions}

We have found that any low $x$ cross section for a process of type $p_0 H \rightarrow p_1 p_2 X$, where $H$ is a hadron and $X$ remnants are not measured, can be rewritten into an infinite twist TMD cross section. Similarly, any low $x$ exclusive cross section for a process of type $p_0 H \rightarrow p_1 p_2 H^\prime$, where $H$ (resp. $H^\prime$) is an incoming (resp. outgoing) hadron, can be rewritten into an infinite twist GTMD cross section. Even though we have restricted ourselves to the case where we have one particle in the initial state and two particles in the final state, all the steps of our study can be applied for processes with more than two particles in the final state, or more than one particle in the initial state. Thus, we bridged one of the main gaps between low $x$ and moderate $x$ formulations of perturbative QCD: the apparent difference between the involved non-perturbative matrix elements. 

We have also given a new interpretation of saturation in the low $x$ regime and distinguished three types of saturation: evolutional, kinematic and genuine. In principle, each type of saturation can  be studied separately from the others
and there are easy ways to distinguish them. For example, studying high $\left|\boldsymbol{k}\right|$ processes on dense targets and on dilute targets would probe genuine saturation alone. On the other hand, small $\left|\boldsymbol{k}\right|$ observables would be probes of both genuine and kinematic saturation on dense targets, and of kinematic saturation alone on dilute targets. 

A study of angular correlations can be performed as in \citep{Dumitru:2016jku}, in the whole kinematic range in $\left| \boldsymbol{k} \right|$ using our results and a tensorial decomposition of the involved TMD distributions. It would be very insightful in future studies to focus on subeikonal corrections to low $x$ physics and try to match a TMD framework, similarly to what was performed in this article.

Finally, the most powerful feature of our formulation in terms of standard parton distributions is the possibility to resum easily logarithms of $Q$ and $|\boldsymbol{k}|$ using the known evolution equations for TMD distributions and Sudakov resummations. This could help solve the observed negativity issues for low $x$ cross sections (see for example~\cite{Stasto:2013cha, Stasto:2014sea, Watanabe:2015tja, Ducloue:2016shw}). Indeed the resummation of colinear logarithms via a  similarly colinearly-improved low $x$ evolution equation~\cite{Iancu:2015vea, Iancu:2015joa, Hatta:2016ujq} is one of the most efficient tools to deal with the issue.
A complementary approach to the improved-JIMWLK evolution would be to first apply the $Y_c$ evolution using the regular JIMWLK equation, then rewrite the evolved observables in terms of TMD distributions as was done in this article, and finally resum logarithms of the hard scale and Sudakov logarithms using standard TMD methods. For the leading twist TMD operators, one could also use the evolution equations derived in~\cite{Balitsky:2015qba, Balitsky:2016dgz}.

\section*{Acknowledgments}
RB thanks P.~Taels and R.~Venugopalan for stimulating discussions. The work of TA is supported by Grant No. 2017/26/M/ST2/01074 of the National Science Centre, Poland. The work of RB is supported by the U.S. Department of Energy, Office of Nuclear Physics, under Contracts No. DE-SC0012704 and by an LDRD grant from Brookhaven Science Associates.

\appendix



\bibliographystyle{unsrt}

\bibliography{MasterBibtex}

\begin{thebibliography}{10}

\bibitem{Balitsky:1995ub}
I.~Balitsky.
\newblock {Operator expansion for high-energy scattering}.
\newblock {\em Nucl. Phys.}, B463:99--160, 1996.

\bibitem{Balitsky:1998kc}
I.~Balitsky.
\newblock {Factorization for high-energy scattering}.
\newblock {\em Phys. Rev. Lett.}, 81:2024--2027, 1998.

\bibitem{Balitsky:1998ya}
Ian Balitsky.
\newblock {Factorization and high-energy effective action}.
\newblock {\em Phys. Rev.}, D60:014020, 1999.

\bibitem{Altinoluk:2014oxa}
Tolga Altinoluk, Nestor Armesto, Guillaume Beuf, Mauricio Martinez, and
  Carlos~A. Salgado.
\newblock {Next-to-eikonal corrections in the CGC: gluon production and spin
  asymmetries in pA collisions}.
\newblock {\em JHEP}, 07:068, 2014.

\bibitem{Altinoluk:2015gia}
Tolga Altinoluk, Nestor Armesto, Guillaume Beuf, and Alexis Moscoso.
\newblock {Next-to-next-to-eikonal corrections in the CGC}.
\newblock {\em JHEP}, 01:114, 2016.

\bibitem{Altinoluk:2015xuy}
Tolga Altinoluk and Adrian Dumitru.
\newblock {Particle production in high-energy collisions beyond the shockwave
  limit}.
\newblock {\em Phys. Rev.}, D94(7):074032, 2016.

\bibitem{Agostini:2019avp}
Pedro Agostini, Tolga Altinoluk, and Nestor Armesto.
\newblock {Non-eikonal corrections to multi-particle production in the Color
  Glass Condensate}.
\newblock 2019.

\bibitem{Balitsky:2015qba}
I.~Balitsky and A.~Tarasov.
\newblock {Rapidity evolution of gluon TMD from low to moderate x}.
\newblock {\em JHEP}, 10:017, 2015.

\bibitem{Balitsky:2016dgz}
I.~Balitsky and A.~Tarasov.
\newblock {Gluon TMD in particle production from low to moderate x}.
\newblock {\em JHEP}, 06:164, 2016.

\bibitem{Balitsky:2017flc}
I.~Balitsky and A.~Tarasov.
\newblock {Higher-twist corrections to gluon TMD factorization}.
\newblock {\em JHEP}, 07:095, 2017.

\bibitem{Balitsky:2017gis}
I.~Balitsky and A.~Tarasov.
\newblock {Power corrections to TMD factorization for Z-boson production}.
\newblock {\em JHEP}, 05:150, 2018.

\bibitem{Chirilli:2018kkw}
Giovanni~Antonio Chirilli.
\newblock {Sub-eikonal corrections to scattering amplitudes at high energy}.
\newblock {\em JHEP}, 01:118, 2019.

\bibitem{Kovchegov:2015pbl}
Yuri~V. Kovchegov, Daniel Pitonyak, and Matthew~D. Sievert.
\newblock {Helicity Evolution at Small-x}.
\newblock {\em JHEP}, 01:072, 2016.
\newblock [Erratum: JHEP10,148(2016)].

\bibitem{Kovchegov:2016zex}
Yuri~V. Kovchegov, Daniel Pitonyak, and Matthew~D. Sievert.
\newblock {Helicity Evolution at Small $x$: Flavor Singlet and Non-Singlet
  Observables}.
\newblock {\em Phys. Rev.}, D95(1):014033, 2017.

\bibitem{Dominguez:2010xd}
Fabio Dominguez, Bo-Wen Xiao, and Feng Yuan.
\newblock {$k_t$-factorization for Hard Processes in Nuclei}.
\newblock {\em Phys. Rev. Lett.}, 106:022301, 2011.

\bibitem{Dominguez:2011wm}
Fabio Dominguez, Cyrille Marquet, Bo-Wen Xiao, and Feng Yuan.
\newblock {Universality of Unintegrated Gluon Distributions at small x}.
\newblock {\em Phys. Rev.}, D83:105005, 2011.

\bibitem{Altinoluk:2018uax}
Tolga Altinoluk, Nestor Armesto, Alex Kovner, Michael Lublinsky, and Elena
  Petreska.
\newblock {Soft photon and two hard jets forward production in proton-nucleus
  collisions}.
\newblock {\em JHEP}, 04:063, 2018.

\bibitem{Altinoluk:2018byz}
Tolga Altinoluk, Renaud Boussarie, Cyrille Marquet, and Pieter Taels.
\newblock {TMD factorization for dijets + photon production from the
  dilute-dense CGC framework}.
\newblock 2018.

\bibitem{Altinoluk:2019fui}
Tolga Altinoluk, Renaud Boussarie, and Piotr Kotko.
\newblock {Interplay of the CGC and TMD frameworks to all orders in kinematic
  twist}.
\newblock 2019.

\bibitem{Metz:2011wb}
Andreas Metz and Jian Zhou.
\newblock {Distribution of linearly polarized gluons inside a large nucleus}.
\newblock {\em Phys. Rev.}, D84:051503, 2011.

\bibitem{Akcakaya:2012si}
Emin Akcakaya, Andreas Schäfer, and Jian Zhou.
\newblock {Azimuthal asymmetries for quark pair production in pA collisions}.
\newblock {\em Phys. Rev.}, D87(5):054010, 2013.

\bibitem{Dumitru:2016jku}
Adrian Dumitru and Vladimir Skokov.
\newblock {$cos(4φ$) azimuthal anisotropy in small-$x$ DIS dijet production
  beyond the leading power TMD limit}.
\newblock {\em Phys. Rev.}, D94(1):014030, 2016.

\bibitem{Marquet:2016cgx}
C.~Marquet, E.~Petreska, and C.~Roiesnel.
\newblock {Transverse-momentum-dependent gluon distributions from JIMWLK
  evolution}.
\newblock {\em JHEP}, 10:065, 2016.

\bibitem{Boer:2017xpy}
Daniel Boer, Piet~J. Mulders, Jian Zhou, and Ya-jin Zhou.
\newblock {Suppression of maximal linear gluon polarization in angular
  asymmetries}.
\newblock {\em JHEP}, 10:196, 2017.

\bibitem{Marquet:2017xwy}
Cyrille Marquet, Claude Roiesnel, and Pieter Taels.
\newblock {Linearly polarized small-$x$ gluons in forward heavy-quark pair
  production}.
\newblock {\em Phys. Rev.}, D97(1):014004, 2018.

\bibitem{Petreska:2018cbf}
Elena Petreska.
\newblock {TMD gluon distributions at small x in the CGC theory}.
\newblock {\em Int. J. Mod. Phys.}, E27(05):1830003, 2018.

\bibitem{Hatta:2016dxp}
Yoshitaka Hatta, Bo-Wen Xiao, and Feng Yuan.
\newblock {Probing the Small- x Gluon Tomography in Correlated Hard Diffractive
  Dijet Production in Deep Inelastic Scattering}.
\newblock {\em Phys. Rev. Lett.}, 116(20):202301, 2016.

\bibitem{Boussarie:2018zwg}
Renaud Boussarie, Yoshitaka Hatta, Bo-Wen Xiao, and Feng Yuan.
\newblock {Probing the Weizsäcker-Williams gluon Wigner distribution in $pp$
  collisions}.
\newblock {\em Phys. Rev.}, D98(7):074015, 2018.

\bibitem{Hatta:2017cte}
Yoshitaka Hatta, Bo-Wen Xiao, and Feng Yuan.
\newblock {Gluon Tomography from Deeply Virtual Compton Scattering at Small-x}.
\newblock {\em Phys. Rev.}, D95(11):114026, 2017.

\bibitem{Belitsky:2003nz}
Andrei~V. Belitsky, Xiang-dong Ji, and Feng Yuan.
\newblock {Quark imaging in the proton via quantum phase space distributions}.
\newblock {\em Phys. Rev.}, D69:074014, 2004.

\bibitem{Lorce:2011kd}
C.~Lorce and B.~Pasquini.
\newblock {Quark Wigner Distributions and Orbital Angular Momentum}.
\newblock {\em Phys. Rev.}, D84:014015, 2011.

\bibitem{Kuraev:1977fs}
E.~A. Kuraev, L.~N. Lipatov, and Victor~S. Fadin.
\newblock {The Pomeranchuk Singularity in Nonabelian Gauge Theories}.
\newblock {\em Sov. Phys. JETP}, 45:199--204, 1977.
\newblock [Zh. Eksp. Teor. Fiz.72,377(1977)].

\bibitem{Balitsky:1978ic}
I.~I. Balitsky and L.~N. Lipatov.
\newblock {The Pomeranchuk Singularity in Quantum Chromodynamics}.
\newblock {\em Sov. J. Nucl. Phys.}, 28:822--829, 1978.
\newblock [Yad. Fiz.28,1597(1978)].

\bibitem{Mueller:1989st}
Alfred~H. Mueller.
\newblock {Small x Behavior and Parton Saturation: A QCD Model}.
\newblock {\em Nucl. Phys.}, B335:115--137, 1990.

\bibitem{Mueller:1993rr}
Alfred~H. Mueller.
\newblock {Soft gluons in the infinite momentum wave function and the BFKL
  pomeron}.
\newblock {\em Nucl. Phys.}, B415:373--385, 1994.

\bibitem{Mueller:1994gb}
Alfred~H. Mueller.
\newblock {Unitarity and the BFKL pomeron}.
\newblock {\em Nucl. Phys.}, B437:107--126, 1995.

\bibitem{McLerran:1993ni}
Larry~D. McLerran and Raju Venugopalan.
\newblock {Computing quark and gluon distribution functions for very large
  nuclei}.
\newblock {\em Phys. Rev.}, D49:2233--2241, 1994.

\bibitem{McLerran:1993ka}
Larry~D. McLerran and Raju Venugopalan.
\newblock {Gluon distribution functions for very large nuclei at small
  transverse momentum}.
\newblock {\em Phys. Rev.}, D49:3352--3355, 1994.

\bibitem{McLerran:1994vd}
Larry~D. McLerran and Raju Venugopalan.
\newblock {Green's functions in the color field of a large nucleus}.
\newblock {\em Phys. Rev.}, D50:2225--2233, 1994.

\bibitem{Gelis:2010nm}
Francois Gelis, Edmond Iancu, Jamal Jalilian-Marian, and Raju Venugopalan.
\newblock {The Color Glass Condensate}.
\newblock {\em Ann. Rev. Nucl. Part. Sci.}, 60:463--489, 2010.

\bibitem{JalilianMarian:1997jx}
Jamal Jalilian-Marian, Alex Kovner, Andrei Leonidov, and Heribert Weigert.
\newblock {The BFKL equation from the Wilson renormalization group}.
\newblock {\em Nucl. Phys.}, B504:415--431, 1997.

\bibitem{JalilianMarian:1997gr}
Jamal Jalilian-Marian, Alex Kovner, Andrei Leonidov, and Heribert Weigert.
\newblock {The Wilson renormalization group for low x physics: Towards the high
  density regime}.
\newblock {\em Phys. Rev.}, D59:014014, 1998.

\bibitem{JalilianMarian:1997dw}
Jamal Jalilian-Marian, Alex Kovner, and Heribert Weigert.
\newblock {The Wilson renormalization group for low x physics: Gluon evolution
  at finite parton density}.
\newblock {\em Phys. Rev.}, D59:014015, 1998.

\bibitem{Kovner:1999bj}
Alex Kovner and J.~Guilherme Milhano.
\newblock {Vector potential versus color charge density in low x evolution}.
\newblock {\em Phys. Rev.}, D61:014012, 2000.

\bibitem{Kovner:2000pt}
Alex Kovner, J.~Guilherme Milhano, and Heribert Weigert.
\newblock {Relating different approaches to nonlinear QCD evolution at finite
  gluon density}.
\newblock {\em Phys. Rev.}, D62:114005, 2000.

\bibitem{Weigert:2000gi}
Heribert Weigert.
\newblock {Unitarity at small Bjorken x}.
\newblock {\em Nucl. Phys.}, A703:823--860, 2002.

\bibitem{Iancu:2000hn}
Edmond Iancu, Andrei Leonidov, and Larry~D. McLerran.
\newblock {Nonlinear gluon evolution in the color glass condensate. 1.}
\newblock {\em Nucl. Phys.}, A692:583--645, 2001.

\bibitem{Ferreiro:2001qy}
Elena Ferreiro, Edmond Iancu, Andrei Leonidov, and Larry McLerran.
\newblock {Nonlinear gluon evolution in the color glass condensate. 2.}
\newblock {\em Nucl. Phys.}, A703:489--538, 2002.

\bibitem{Kovchegov:1999yj}
Yuri~V. Kovchegov.
\newblock {Small x F(2) structure function of a nucleus including multiple
  pomeron exchanges}.
\newblock {\em Phys. Rev.}, D60:034008, 1999.

\bibitem{Gribov:1984tu}
L.~V. Gribov, E.~M. Levin, and M.~G. Ryskin.
\newblock {Semihard Processes in QCD}.
\newblock {\em Phys. Rept.}, 100:1--150, 1983.

\bibitem{Dumitru:2005gt}
Adrian Dumitru, Arata Hayashigaki, and Jamal Jalilian-Marian.
\newblock {The Color glass condensate and hadron production in the forward
  region}.
\newblock {\em Nucl. Phys.}, A765:464--482, 2006.

\bibitem{Altinoluk:2011qy}
T.~Altinoluk and A.~Kovner.
\newblock {Particle Production at High Energy and Large Transverse Momentum -
  'The Hybrid Formalism' Revisited}.
\newblock {\em Phys. Rev.}, D83:105004, 2011.

\bibitem{Chirilli:2011km}
Giovanni~A. Chirilli, Bo-Wen Xiao, and Feng Yuan.
\newblock {One-loop Factorization for Inclusive Hadron Production in $pA$
  Collisions in the Saturation Formalism}.
\newblock {\em Phys. Rev. Lett.}, 108:122301, 2012.

\bibitem{Chirilli:2012jd}
Giovanni~A. Chirilli, Bo-Wen Xiao, and Feng Yuan.
\newblock {Inclusive Hadron Productions in pA Collisions}.
\newblock {\em Phys. Rev.}, D86:054005, 2012.

\bibitem{Stasto:2013cha}
Anna~M. Stasto, Bo-Wen Xiao, and David Zaslavsky.
\newblock {Towards the Test of Saturation Physics Beyond Leading Logarithm}.
\newblock {\em Phys. Rev. Lett.}, 112(1):012302, 2014.

\bibitem{Stasto:2014sea}
Anna~M. Stasto, Bo-Wen Xiao, Feng Yuan, and David Zaslavsky.
\newblock {Matching collinear and small $x$ factorization calculations for
  inclusive hadron production in $pA$ collisions}.
\newblock {\em Phys. Rev.}, D90(1):014047, 2014.

\bibitem{Altinoluk:2014eka}
Tolga Altinoluk, Nestor Armesto, Guillaume Beuf, Alex Kovner, and Michael
  Lublinsky.
\newblock {Single-inclusive particle production in proton-nucleus collisions at
  next-to-leading order in the hybrid formalism}.
\newblock {\em Phys. Rev.}, D91(9):094016, 2015.

\bibitem{Watanabe:2015tja}
Kazuhiro Watanabe, Bo-Wen Xiao, Feng Yuan, and David Zaslavsky.
\newblock {Implementing the exact kinematical constraint in the saturation
  formalism}.
\newblock {\em Phys. Rev.}, D92(3):034026, 2015.

\bibitem{Ducloue:2016shw}
B.~Ducloue, T.~Lappi, and Y.~Zhu.
\newblock {Single inclusive forward hadron production at next-to-leading
  order}.
\newblock {\em Phys. Rev.}, D93(11):114016, 2016.

\bibitem{Iancu:2016vyg}
E.~Iancu, A.~H. Mueller, and D.~N. Triantafyllopoulos.
\newblock {CGC factorization for forward particle production in proton-nucleus
  collisions at next-to-leading order}.
\newblock {\em JHEP}, 12:041, 2016.

\bibitem{Ducloue:2017dit}
B.~Ducloue, E.~Iancu, T.~Lappi, A.~H. Mueller, G.~Soyez, D.~N.
  Triantafyllopoulos, and Y.~Zhu.
\newblock {Use of a running coupling in the NLO calculation of forward hadron
  production}.
\newblock {\em Phys. Rev.}, D97(5):054020, 2018.

\bibitem{Ducloue:2017mpb}
B.~Ducloue, T.~Lappi, and Y.~Zhu.
\newblock {Implementation of NLO high energy factorization in single inclusive
  forward hadron production}.
\newblock {\em Phys. Rev.}, D95(11):114007, 2017.

\bibitem{Kotko:2015ura}
P.~Kotko, K.~Kutak, C.~Marquet, E.~Petreska, S.~Sapeta, and A.~van Hameren.
\newblock {Improved TMD factorization for forward dijet production in
  dilute-dense hadronic collisions}.
\newblock {\em JHEP}, 09:106, 2015.

\bibitem{vanHameren:2016ftb}
A.~van Hameren, P.~Kotko, K.~Kutak, C.~Marquet, E.~Petreska, and S.~Sapeta.
\newblock {Forward di-jet production in p+Pb collisions in the small-x improved
  TMD factorization framework}.
\newblock {\em JHEP}, 12:034, 2016.

\bibitem{Mueller:1986ey}
Alfred~H. Mueller and H.~Navelet.
\newblock {An Inclusive Minijet Cross-Section and the Bare Pomeron in QCD}.
\newblock {\em Nucl. Phys.}, B282:727--744, 1987.

\bibitem{Iancu:2015vea}
E.~Iancu, J.~D. Madrigal, A.~H. Mueller, G.~Soyez, and D.~N.
  Triantafyllopoulos.
\newblock {Resumming double logarithms in the QCD evolution of color dipoles}.
\newblock {\em Phys. Lett.}, B744:293--302, 2015.

\bibitem{Iancu:2015joa}
E.~Iancu, J.~D. Madrigal, A.~H. Mueller, G.~Soyez, and D.~N.
  Triantafyllopoulos.
\newblock {Collinearly-improved BK evolution meets the HERA data}.
\newblock {\em Phys. Lett.}, B750:643--652, 2015.

\bibitem{Hatta:2016ujq}
Yoshitaka Hatta and Edmond Iancu.
\newblock {Collinearly improved JIMWLK evolution in Langevin form}.
\newblock {\em JHEP}, 08:083, 2016.

\end{thebibliography}

\end{document}